\documentclass[structabstract]{aa}  
\usepackage{graphicx}
\usepackage{tabularx, array}
\usepackage{float}
\usepackage{txfonts}
\usepackage{natbib}
\usepackage{longtable}
\usepackage{rotating}
\usepackage{lscape}
\usepackage{appendix}
\bibpunct[; ]{(}{)}{;}{a}{}{,}
\begin{document}
\title{The Infrared Telescope Facility (IRTF) spectral 
library: spectral diagnostics for cool stars}


\author{M. Cesetti\inst{1,2}
        \and
        A. Pizzella\inst{1,2}
        \and
        V. D. Ivanov\inst{3}
        \and
        L. Morelli\inst{1,2}
        \and
        E. M. Corsini\inst{1,2}
        \and
        E. Dalla Bont\`a\inst{1,2}
        }
        
\offprints{M. Cesetti}

\institute{Dipartimento di Fisica e Astronomia ``G. Galilei'', 
           Universit\`a di Padova,
           vicolo dell'Osservatorio 3, I-35122 Padova, Italy\\
          \email{mary.cesetti@unipd.it}\\
          \and
          INAF--Osservatorio Astronomico di Padova, vicolo
	  dell'Osservatorio 5, I-35122 Padova, Italy\\         
          \and    
          European Southern Observatory, Ave. Alonso de
          C\`ordova 3107, Casilla 19, Santiago 19001, Chile
           }

\date{}


\abstract
{The near-infrared (NIR) wavelength range offers some unique spectral
features, and it is less prone to the extinction than the optical one.
Recently, the first flux calibrated NIR library of cool stars from the
NASA Infrared Telescope Facility (IRTF) have become available, and it
has not been fully exploited yet.}
{We want to develop spectroscopic diagnostics for stellar physical
parameters based on features in the wavelength range $1-5\,\mu$m. In
this work we test the technique in the $I$ and $K$ bands. The study of
the $Y$, $J$, $H$, and $L$ bands will be presented in the following
paper.}
{An objective method for semi-empirical definition of spectral
features sensitive to various physical parameters is applied to the
spectra. It is based on {\it sensitivity map}--i.e., derivative of
the flux in the spectra with respect to the stellar parameters at a
fixed wavelength. New optimized indices are defined and their
equivalent widths (EWs) are measured.}
{The method is applied in the $I$- and $K$-band windows of the IRTF
stellar spectra to verify the new technique by comparing the results
with the known behavior of well-studied spectral features. A number of
sensitive features to the effective temperature and surface gravity
are re-identified or newly identified clearly showing the reliability
of the {\it sensitivity map}~analysis.}
{The {\it sensitivity map}~allows to identify the best bandpass limits
for the line and nearby continuum. It reliably predicts the trends of
spectral features with respect to a given physical parameter but not
their absolute strengths. Line blends are easy to recognize when
blended features have different behavior with respect to some physical
stellar parameter. The use of sensitivity map is therefore
complementary to the use of indices. We give the EWs of the new
indices measured for the IRTF star sample. This new and homogeneous
set of EWs will be useful for stellar population synthesis models and
can be used to get element-by-element abundances for unresolved
stellar population studies in galaxies.}


\titlerunning{Near-infrared spectral line diagnostics for cool stars} 

\authorrunning{Cesetti et al.}

\maketitle


\section{Introduction}

The interpretation of the spectral absorption features in the
integrated light of galaxies requires the understanding of the
behavior of these features in the spectra of the stars as function of
their effective temperature $T_{\rm eff}$, surface gravity
$\log\,(g)$, and metal abundance (usually parametrized with the Iron
abundance [Fe/H] and the ratio [$\alpha$/Fe] of alpha-elements to
Iron). This explains why some major observational and theoretical
efforts are invested into the development of extensive stellar
libraries for population synthesis e.g.,
\citep{Bur84,Wor94,Tra98,Tri95,Kor05}. The knowledge derived from the
analysis of the stellar spectra is promptly transferred to the galaxy
integrated light because $T_{\rm eff}$ and $\log\,(g)$ are related to
the age of the dominant stellar population, and [Fe/H] is related to
the overall galactic chemical abundance.

Usually, first the sensitivity of the spectral absorption features to
the physical parameters of stars is empirically investigated, and then
stellar photosphere models are adopted to explain the behavior of the
features within the theory of line formation. The former step is
typically limited to looking for correlations between of the line
strengths and $T_{\rm eff}$, $\log\,(g)$, or [Fe/H]. A notable
exception is the work by \citet[][see their Tables\,2 and 3]{Wor94}
who determined the derivatives of these relations and adopted them as
{\it sensitivity indices} for individual lines in order to compare the
relative sensitivity of various features to the physical parameters of
the stellar population.

Prompted by this idea, we attempt here to develop it further to its
logical conclusion by defining the {\it sensitivity map}, that is
a continuous derivative across the stellar spectra of stars for
different $T_{\rm eff}$, $\log\,(g)$, or [Fe/H] spanning the full
range of parameters for a given wavelength.

This new method allowed us to characterize the behavior of the
individual spectral features and it is an useful and objective
tool to define at the best the spectral indices more sensible 
to physical stellar parameters. Moreover, it accounts
for the variable sensitivity of individual lines. It was
developed in order to explore near-infrared (NIR) spectral
ranges, which have never been used for stellar population analysis
before (i.e., $1-1.8\,\mu$m). First we applied it to some
well-understood features, such as the Calcium Triplet (CaT
  hereafter) at 0.85$\,\mu$m and CO band at 2.29$\,\mu$m, to
demonstrate its feasibility. In this paper we also present the
sensitivity map for a number of other NIR indices with respect
to spectral type (SpT hereafter) and surface gravity. In a
forthcoming paper we will extend such an analysis to other
spectral ranges to define new indices, optimized for stellar
population analysis. The new set of NIR indices will be useful
for both stellar and extragalactic astrophysics.

The structure of the paper is as follows. Section\,\ref{sec:IRTF}
gives a short description of the adopted Infrared Telescope Facility
(IRTF) spectral library. Section\,\ref{sec:phys_params} summarizes the
physical parameters of the stars in the library. The method used to
create the sensitivity map is described in
Sect.\,\ref{sec:method}. The absorption features and line indices in
the $I$ band and their use as spectral diagnostics are discussed in
Sects.\,\ref{sec:sp_ind_I} and \ref{sec:SpT_diag_I}, respectively.
Sections\,\ref{sec:sp_ind_K} and \ref{sec:SpT_diag_K} are devoted to
$K$-band indices. Section\,\ref{Sec:limits} 
discusses and summarizes the results.


\section{The IRTF spectral library}
\label{sec:IRTF}

The IRTF spectral library \citep{Cus05, Ray09}, which is the first NIR
library of flux-calibrated stellar spectra and extends the optical
library described in \citet{Gor93}, contains 210 cool stars. The
spectra were taken with the cross-disperser medium-resolution infrared
spectrograph SpeX \citep{Ray03} mounted at the 3.0-meter NASA Infrared
Telescope Facility (IRTF) on Mauna Kea, Hawaii. The observations were
carried out with two different set-ups with a resolving power
$R\,\approx\,2000$ at $0.8-2.4\,\mu$m and $R\,\approx\,2500$ at
$2.4-5\,\mu$m, respectively. Several spectral orders were
simultaneously recorded during a single exposure with significant
wavelength overlap between the adjacent orders making it easier to
preserve the continuum shape. It results to be reliable to within a
few percent, as verified by generating a set of synthetic Two-Micron
All Sky Survey (2MASS) colors from the spectra
\citep{Ray09}. Therefore, the IRTF library allowed us to measure the
strong and broad molecular absorption bands that are common in the
NIR.

Most of the stars were observed in the $0.8-2.5\,\mu$m range, and for
a small fraction of them the wavelength coverage extends up to
5$\,\mu$m. The wide wavelength range permits to connect some
well-studied regions as the $I$ and $K$ bands, with the relatively
unexplored $J$, $H$, and $L$ bands. The sample contains the F, G, K,
M, and L stars, spanning luminosity classes from I to V. It also
includes some AGB, carbon, and S stars. The SpTs were derived from
optical spectra in the framework of the MK classification system
\citep{Ray09}.

The sample is limited to bright stars (Fig.\,\ref{fig:Tefg_TefLk}),
 guaranteeing high signal-to-noise ($S/N$) ratio even in the thermal
 infrared region ($\sim2.3-5\,\mu$m). It is $S/N \ge 100$ across most
 of the wavelength range, with the exception of regions with poor
 atmospheric transmission and with $\lambda\,>4\,\mu$m. The brightness
 limit has a downside. Indeed, most of the sample stars are located
 nearby in the Milky Ways disk, and they have near-solar chemical
 composition. The abundance distribution of the sample stars with
 known [Fe/H] \citep{Cay97} is shown in Fig.\,1(a) of \citet{Ray09},
 and it is representative for the stars in the solar neighborhood
 \citep{Nor04}.


\section{Physical parameters of the sample stars}
\label{sec:phys_params}

The sample stars span a wide range in SpTs and luminosity
classes. The photospheric data for the sample stars are taken from
literature. This is an inhomogeneous compilation, and in an effort to
homogenize the data as much as possible (at least in terms of
abundance estimate methods) we used spectroscopic determinations when
possible. We adopted the metallicities based on narrow-band photometry
when these were the only ones available in the literature. The
$K$-band absolute luminosities were calculated by convolving the
spectra with a filter transmission curve. If $\log(T_{\rm eff})$ was
not available from literature, we estimated it from the SpT-$T_{\rm
eff}$ relation by \citet{Car96}. Part of the sample lacks gravity
determination. The parallaxes were obtained from the HIPPARCOS catalog
\citep{Per97}. All the collected parameters of the IRTF stars are
listed in Table\,\ref{tab:sample_params}.

\begin{figure}
\includegraphics[width=0.5\textwidth]{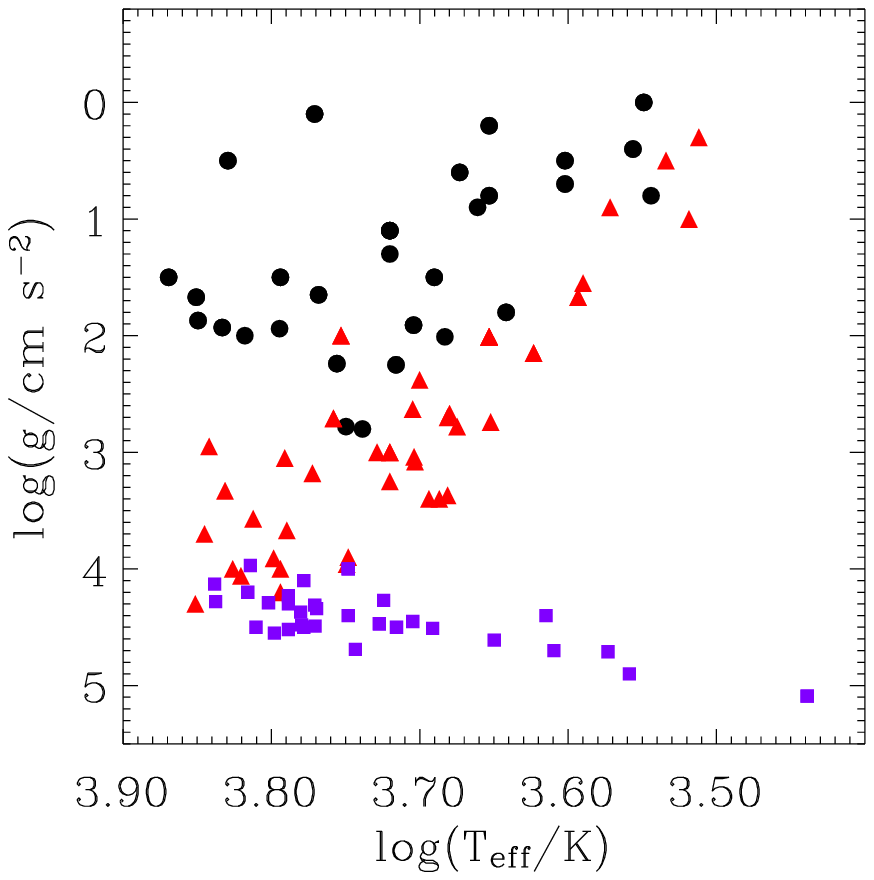}
\includegraphics[width=0.5\textwidth]{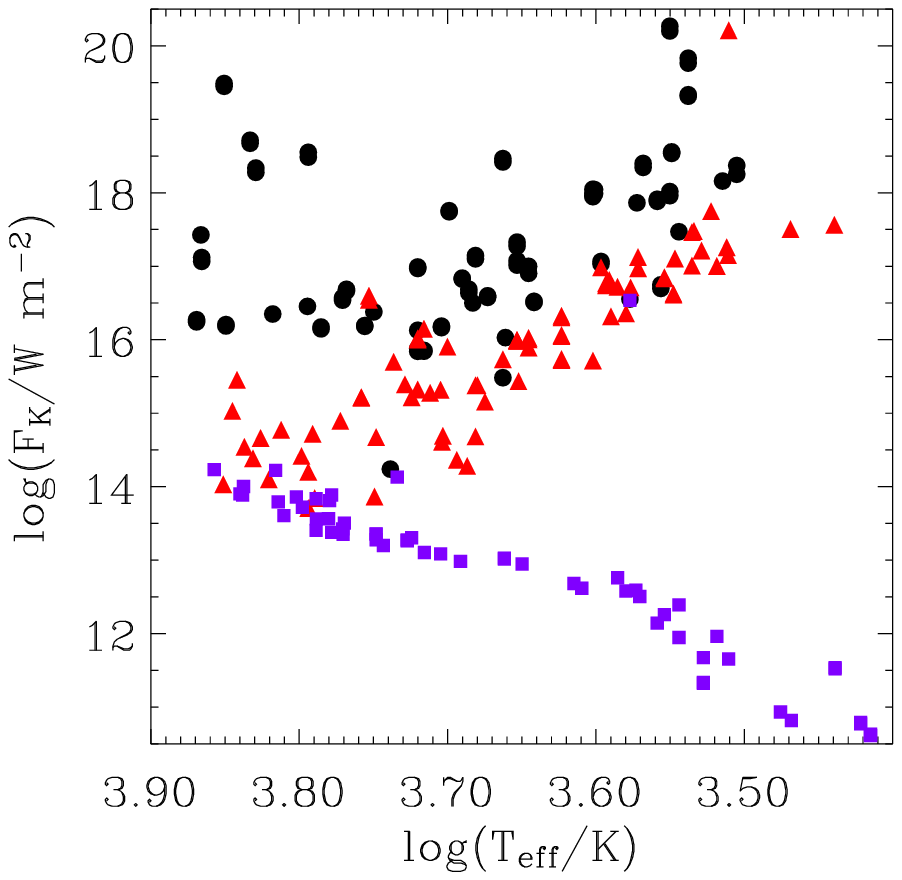}
\caption{Surface gravity $\log\,(g)$ (top panel) and $K$-band flux
$\log(F_{K})$ (bottom panel) as a function of effective
temperature $\log(T_{\rm eff})$ for the sample stars.}
\label{fig:Tefg_TefLk}
\end{figure}

\begin{figure}
\includegraphics[width=9truecm,angle=0]{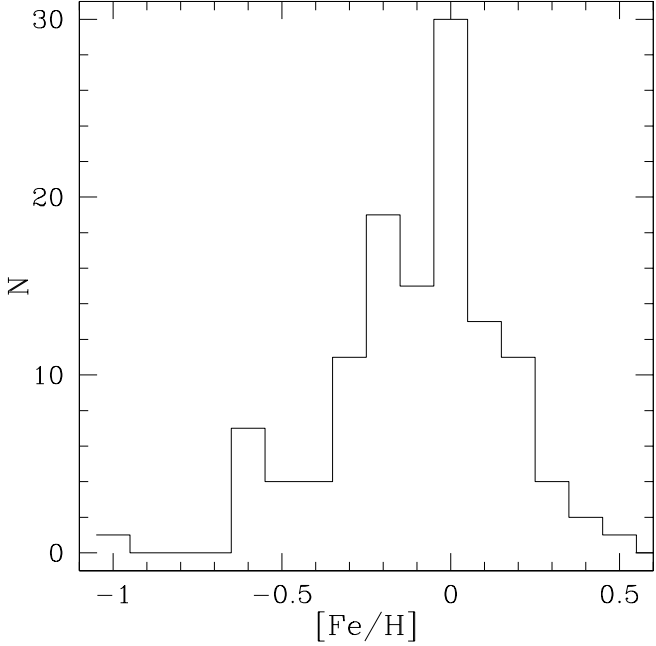}
\caption{Metallicity distribution of the sample stars with
spectroscopic measurements of [Fe/H].}
\label{fig:MetDis}
\end{figure}

Figure\,\ref{fig:Tefg_TefLk} shows how the sample stars populate the
$\log\,(g)-\log(T_{\rm eff})$ and $\log(L_{K})-\log(T_{\rm eff})$
planes. Their metallicity distribution is shown in
Fig.\,\ref{fig:MetDis}. The average [Fe/H] is about $-0.1$ dex with a
dispersion of 0.2 dex, which is typical of the solar neighborhood.

\section{Method}
\label{sec:method}

This section presents the empirical method we used to identify the
most sensitive line-strength indices to a particular stellar physical
parameter, and the procedure to define them. The method is free from
any {\em a priori\/} assumption and it can be applied to any star or
line-strength index. But here it was constrained to the range of SpTs,
luminosity classes, and spectral coverage spanned by the IRTF
library. Our goal is to find proxies for $T_{\rm eff}$, $\log\,(g)$
and [Fe/H], except for that the temperature is replaced by SpT, as it
is more readily available. We numbered the SpT assigning value of 0 to
F0 stars, 1 to G0, 2 to K0, etc, with decimals for the subtypes ({\rm
e.g.}, F5=0.5; G8=1.8; M0.5=3.05).  In the first application described
in this work we considered the spectral ranges that include the CaT
and CO bands that have already been widely used for stellar population
analysis \citep[e.g.,][]{Jon84, Arm91, May97, Iva00, Vaz03, Mar08}.
They provided a test-bed for verifying the method.

The first step was to homogenize the spectra re-binning them with a
step of $3.872\,\times\,10^{-4}\,\mu$m~pixel$^{-1}$ . The supergiants,
giants, and dwarfs were analyzed separately.
After experimenting with absolute flux calibrated spectra and
continuum-normalized spectra we opted the latter because the
normalization lowered the scatter in the final results. The spectra
were normalized to unity at $\lambda\,=\,0.88\,\mu$m and $2.20\,\mu$m
for the analysis of the $I$ and $K$ bands, respectively. Both these
wavelength regions were chosen because they were free of relevant
absorption features.

Next, for each wavelength bin we derived a second-order polynomial
least-square fit of the normalized intensity as a function of SpT. The
fit was performed under the IDL\footnote{The interactive Data Language
is distributed by ITT Visual Information Solutions.}  environment
using a specially developed script.  An independent fit is obtained
for each wavelength bin.  The choice of the second-order is discussed
in Sect.\ref{Sec:limits}. 
We defined as {\it model spectrum}
the combined fitting functions for all wavelengths.  Such a fit was
introduced to provide continuity across the entire parameter space,
otherwise, the sensitivity could be evaluated only for discrete values
of the physical parameter (the SpT in this case) for which stellar
spectra were available.

Finally, we calculated the derivative of the model spectrum with
respect to the SpT by building the so-called {\it sensitivity
map}. 
A spectral
index is sensitive to the SpT if the derivative with respect to the
SpT of the model spectrum at the index central wavelength is different
from the derivative of the surrounding region. The wavelength regions
where such a strong difference is observed are characterized by sharp
features in the sensitivity map and can be easily identified. Cuts of
the fit and sensitivity map for a fixed SpT in the $I$ band are shown
for sample supergiants in Figs.\,\ref{fig:SGiant_fitted} and
\ref{fig:SupGian_SpT}, respectively.

To study the features sensitive to gravity we divided the sample into
four SpT bins (F, G, K, and M; the L stars in the sample have no
gravity measurements and therefore were excluded).  In each bin we
sorted the flux-normalized spectra in a sequence of increasing
$\log\,(g)$. We then followed the same steps as for the SpT, except
for that the fit and derivative were performed along the $\log\,(g)$
axis. Figs.\,\ref{fig:FG_Logg} and \ref{fig:KM_Logg} show the results
of the analysis for fixed surface gravity in the $I$ band for the F
and G stars and the K and M stars, respectively.


\section{Spectral indices in the $I$ band}
\label{sec:sp_ind_I}

\subsection{Main $I$-band spectral features}  
\label{sec:Iband}

Here we shortly summarize the main $I$-band ($0.80-0.90\,\mu$m)
spectral features in later type stars. The spectra of F stars are
dominated by the neutral hydrogen (H{~\sc i}) absorption lines of the
Paschen series. Their strength decreases with increasing
wavelength. The Paschen series becomes weaker progressing from
supergiants through giants to dwarfs, and from F to late-type G
stars. The absorption lines of neutral metals are stronger in G stars
than in F stars, and reach a maximum depth in the spectra of K and M
stars. The lines of ionized metals (the strongest feature is CaT at
$0.86\,\mu$m) weaken towards later SpTs. The molecular absorption
increases in later types, affecting significantly the slope of the
local continuum. In M stars the titanium oxide (TiO) bands are
significant and blend with CaT, which is weaker than in earlier
types. Progressing from M to L stars, the metal oxides (TiO and VO)
are replaced by metal hydrides (CrH at 0.8611\,$\mu$m and FeH at
0.8692\,$\mu$m) as the main molecular species \citep{Ray09}.

The CaT was used in different studies over a wide range of atmospheric
parameters and was applied to both individual stars and integrated
stellar populations in different environments, the calibration is both
empirical \citep[i.e.,][]{Cen02} and theoretical
\citep[i.e.,][]{Du12}. Several definitions for the CaT index exist:
the classical approach consists in establishing a central
bandpass covering the spectral feature, and one or more adjacent
bandpasses to trace the reference level of the local continuum.
\citet[][Cen01 hereafter]{Cen01a} analyzed previous CaT index
definitions and defined a new one, which was specifically
designed to avoid contamination from molecular bands and to cover the
line wings completely. The latter is an important issue, because
the main contributors to the strength of the CaT lines are their
wings, whereas the core is not very sensitive to the 
atmospheric and stellar parameters \citep{Erd91}. The Ca{~\sc ii}
lines are heavily affected by metallicity and gravity: their strengths
increase as metallicity increases and gravity decreases. For a
detailed discussion see Sect.\,2 of Cen01. To summarize, the CaT in
cool stars ($\sim$F7--M0) follows a complex behavior with varying
temperature, metallicity, and gravity.

\subsection{Sensitivity map for the spectral type}
\label{sec:sens_fun1_I}

\begin{figure*}
\includegraphics[width=16truecm,angle=0]{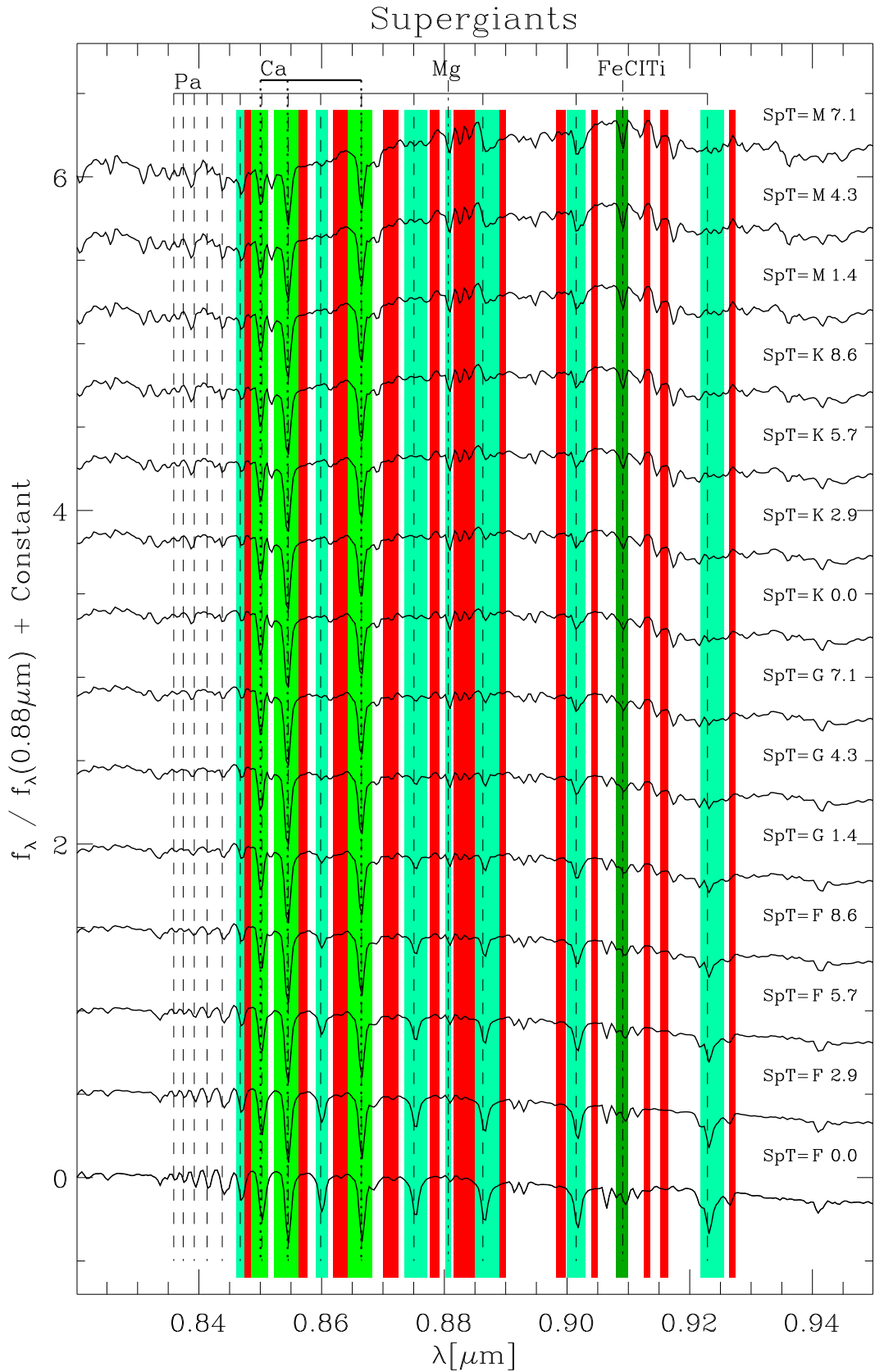}
\caption{$I$-band {\it model spectrum} of supergiant stars obtained by
fitting at each wavelength the flux-normalized (at 0.88$\,\mu$m)
sample spectra along SpTs. The model spectrum for different SpTs is
offset for displaying purposes and the SpT is given. The green and red
regions mark the bandpasses of the newly defined indices and their
adjacent continuum, respectively (see Table\,\ref{tab:index_def_I}). In
particular the light green, green and dark green regions mark the Pa,
Ca, Mg and FeClTe features, respectively. Some relevant absorption
features are marked.}
\label{fig:SGiant_fitted}
\end{figure*}

\begin{figure*}
\includegraphics[width=16truecm,angle=0]{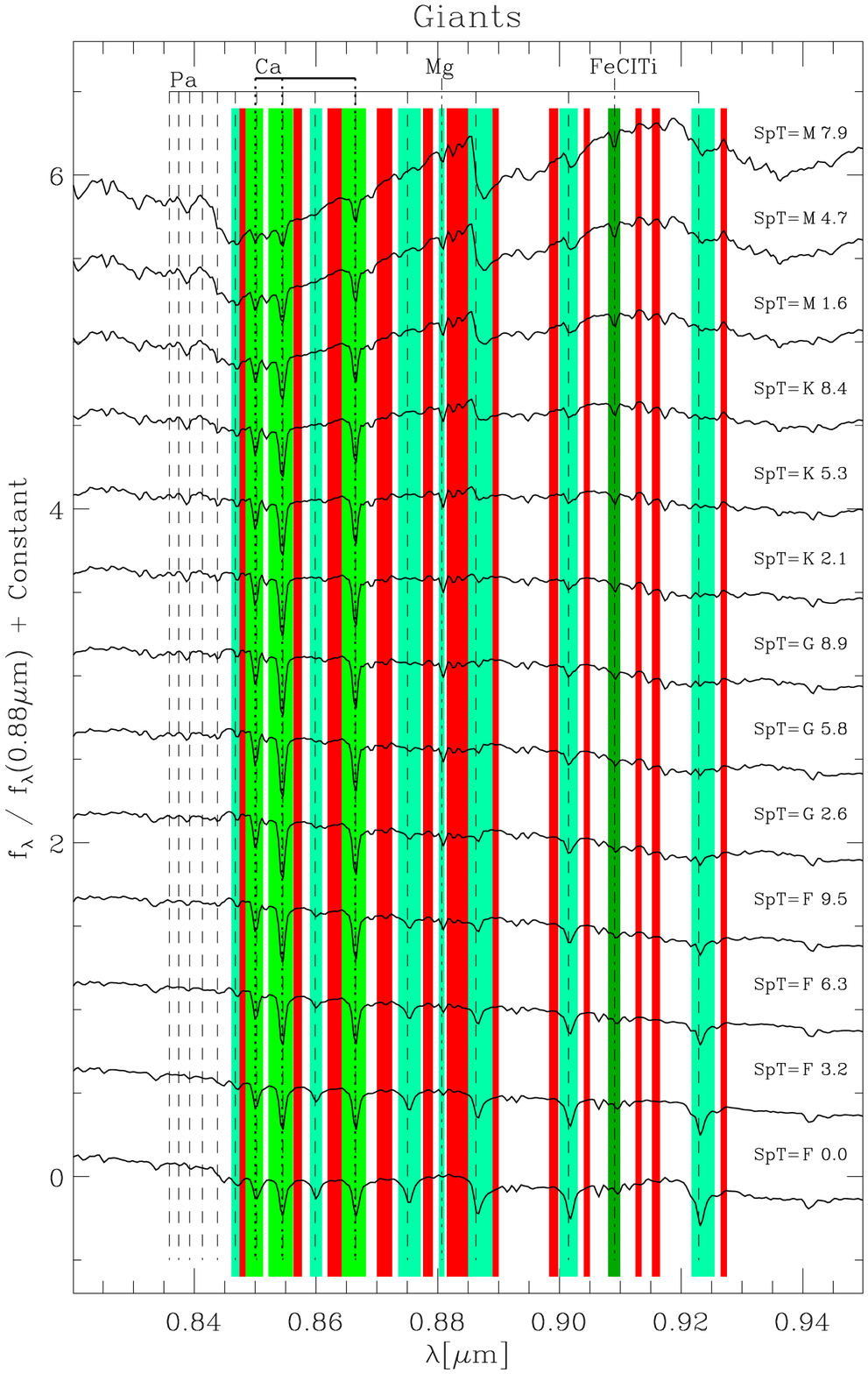}
\caption{As in Fig.\,\ref{fig:SGiant_fitted} but for giant stars.}
\label{fig:Giant_fitted}
\end{figure*}

\begin{figure*}
\includegraphics[width=16truecm,angle=0]{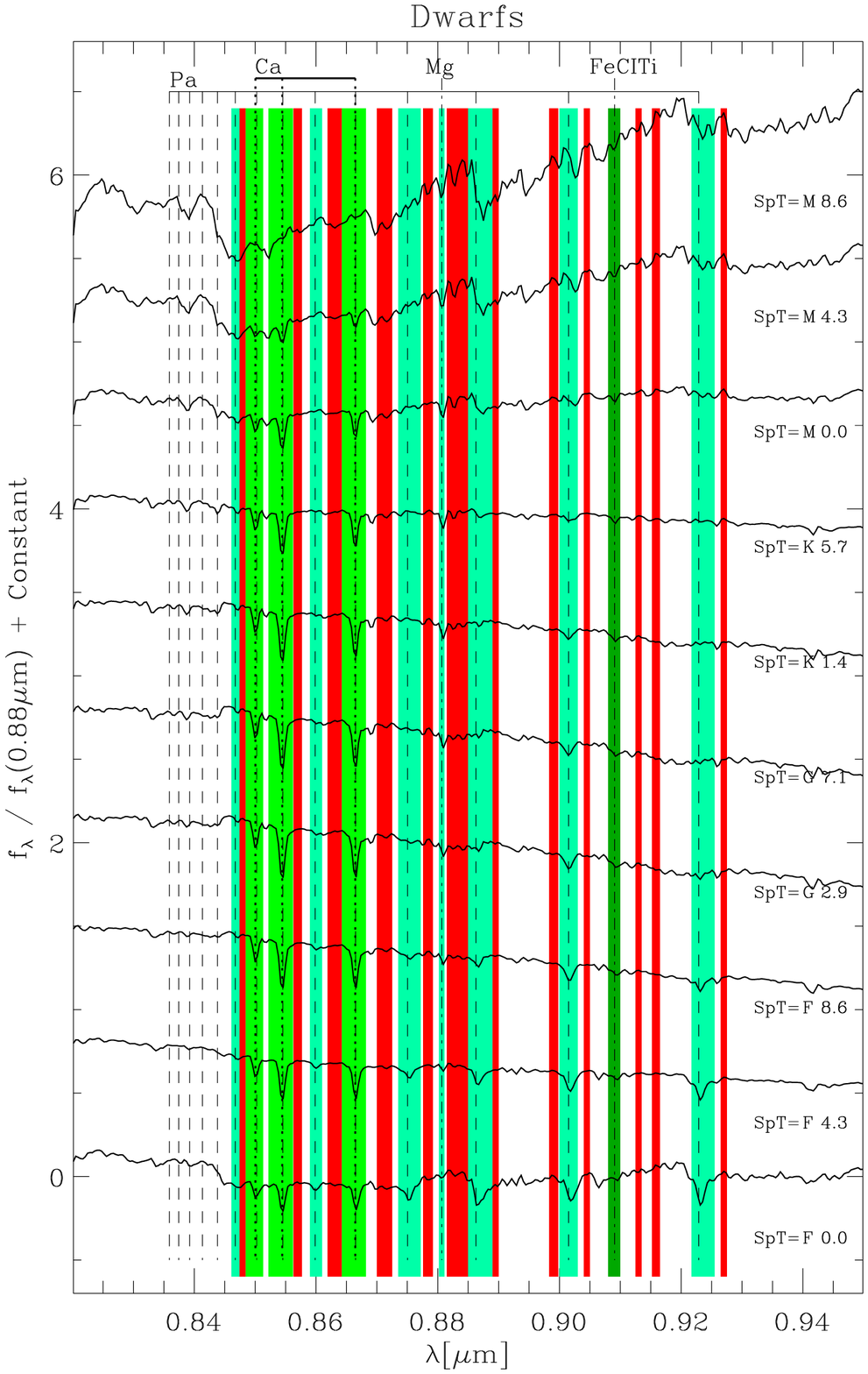}
\caption{As in Fig.\,\ref{fig:SGiant_fitted} but for dwarf stars.}
\label{fig:Dwarf_fitted}
\end{figure*}

\begin{figure*}
\includegraphics[width=16truecm,angle=0]{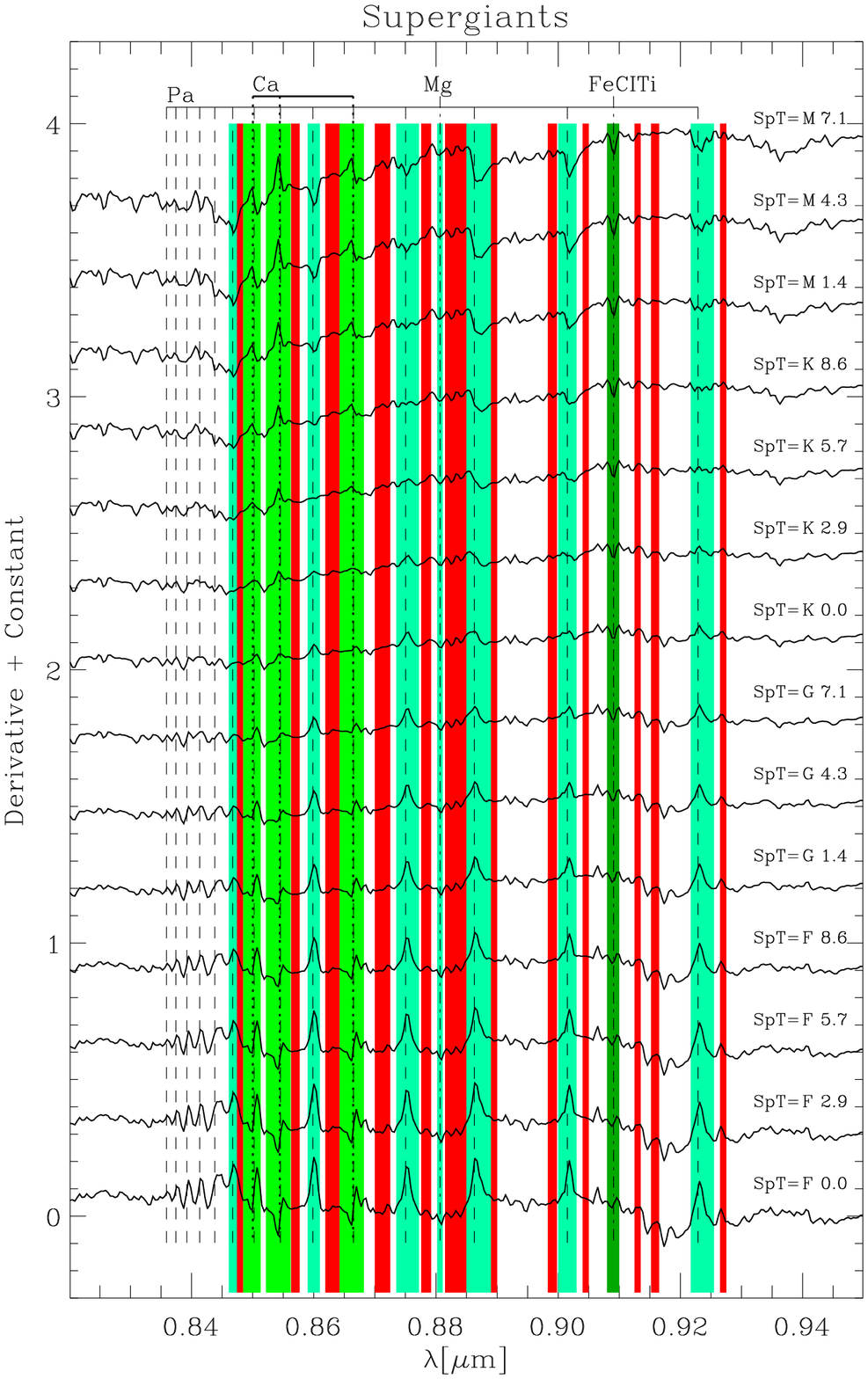}
\caption{$I$-band {\it sensitivity map} for SpT of supergiant stars.
The sensitivity map for different SpTs is offset for displaying
purposes and the SpT is given. Symbols are as in
Fig.\,\ref{fig:SGiant_fitted}.}
\label{fig:SupGian_SpT}
\end{figure*}

\begin{figure*}
\includegraphics[width=16truecm,angle=0]{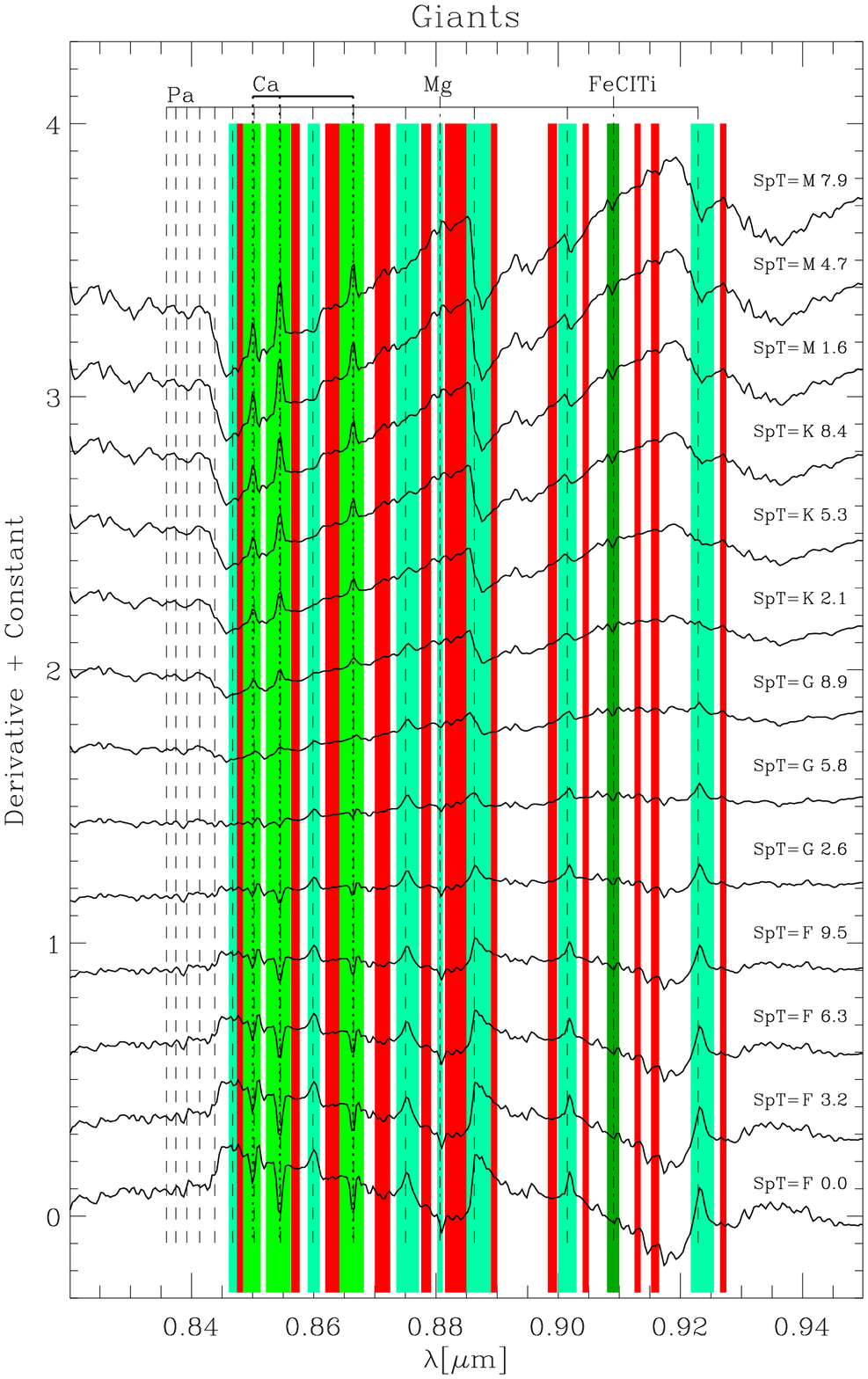}
\caption{As in Fig.\,\ref{fig:SupGian_SpT} but for giant stars.}
\label{fig:Gian_SpT}
\end{figure*}

\begin{figure*}
\includegraphics[width=16truecm,angle=0]{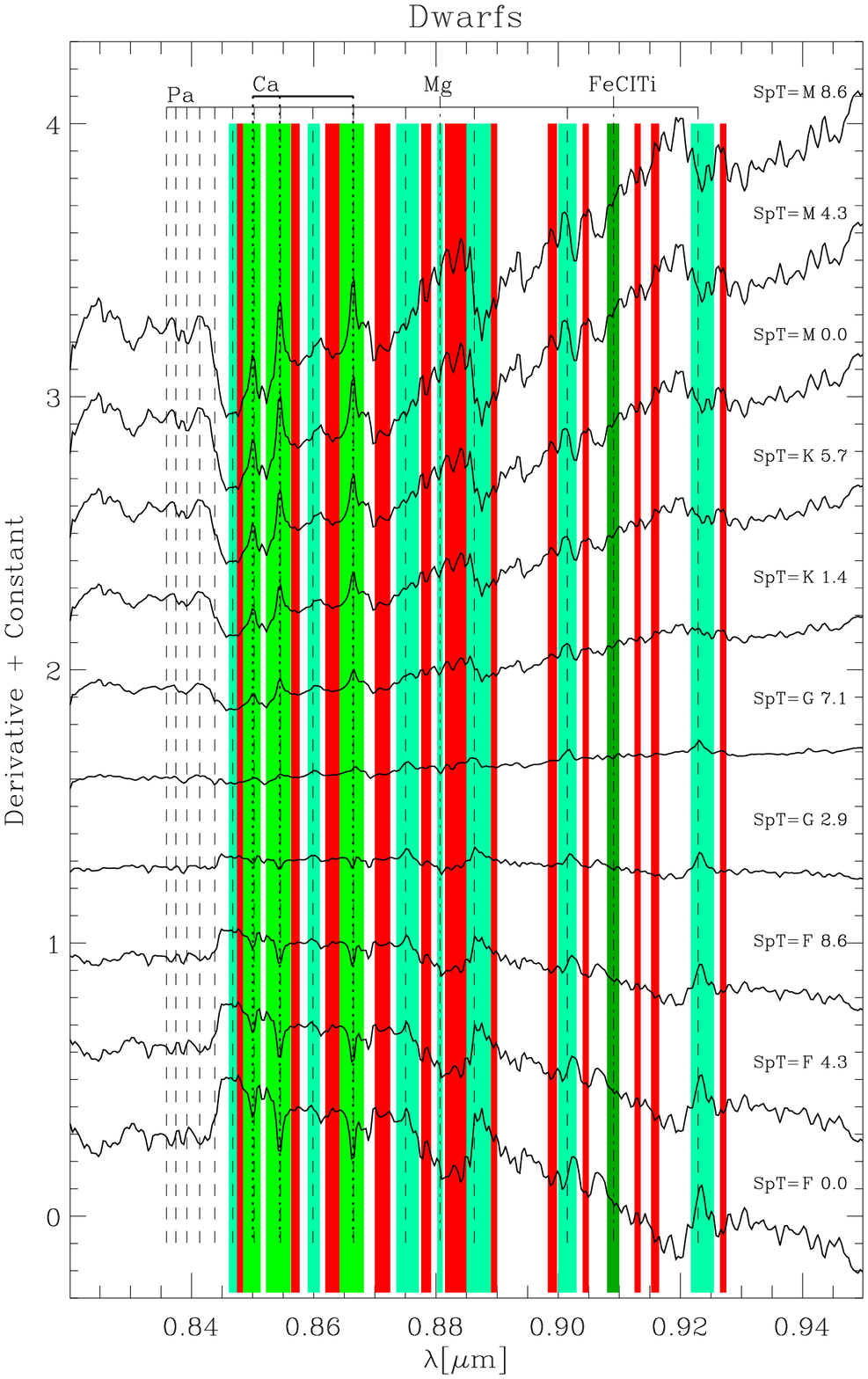}
\caption{As in Fig.\,\ref{fig:SupGian_SpT} but for dwarf stars.}
\label{fig:Dwarf_SpT}
\end{figure*}

The $I$-band model spectrum and sensitivity map for the different SpTs
(i.e., for different $T_{\rm eff}$) are shown in
Figs.\,\ref{fig:SGiant_fitted}, \ref{fig:Giant_fitted}, and
\ref{fig:Dwarf_fitted} and in Figs.\,\ref{fig:SupGian_SpT},
\ref{fig:Gian_SpT}, and \ref{fig:Dwarf_SpT} for supergiant, giant, and
dwarf stars, respectively. The different plots are shown with a
constant vertical offset for display purposes and the position of the
CaT band, Paschen lines, the Mg and the FeClTi band are marked.

The analysis of the supergiant stars can be used to summarize the
features present in the $I$ band (Fig.\,\ref{fig:SupGian_SpT}). The
H{~\sc i} lines from the Paschen series correspond to positive peaks
in the sensitivity map over about the F0--K0 SpT range. Their
equivalent width (EW hereafter) decreases with SpT, because the flux
within the absorption line is growing more rapidly than the flux of
the surrounding continuum. From about K0.0 to about K8.6 the
sensitivity of the Paschen lines to the SpT is negligible, and from
about K8.6 to about M7.1 the sensitivity map shows negative peaks at
the Paschen lines because these features decrease toward later types
over this range of SpTs.  The CaT generally shows just the opposite
trend, but with asymmetric sensitivity map because the features are
contaminated by H{~\sc i}, so there is a superposition of a negative
and a positive peak. Higher-resolution spectroscopy with sufficient
$S/N$ to allow line decomposition is required when CaT is used as a
diagnostic tool. The Magnesium sensitivity map show negative
peaks for stars of all spectral types and luminosity classes.

The samples of giant and dwarf stars show similar behavior
(Figs.\,\ref{fig:Gian_SpT} and \ref{fig:Dwarf_SpT}) but the H{~\sc i}
lines are less-sensitive temperature indicators than in supergiant
stars. As a result, the variation of the CaT comes across clearer,
with symmetric, less contaminated peaks. Some cases of contamination
are still present, e.g., Pa4 at 0.886\,$\mu$m is affected by Ni and Fe
absorption lines. This index should only be used for hotter stars
that show no TiO in their atmospheres.  Finally, the variations of
molecular features are particularly strong in giant stars, and affect
most of the continuum in $I$ band, confirming that they are good
temperature indicators. The FeClTi band shows no variation in dwarf
stars.

\subsection{Sensitivity map for the surface gravity}
\label{sec:sens_fun2_I}

\begin{figure*}
\includegraphics[width=16truecm,angle=0]{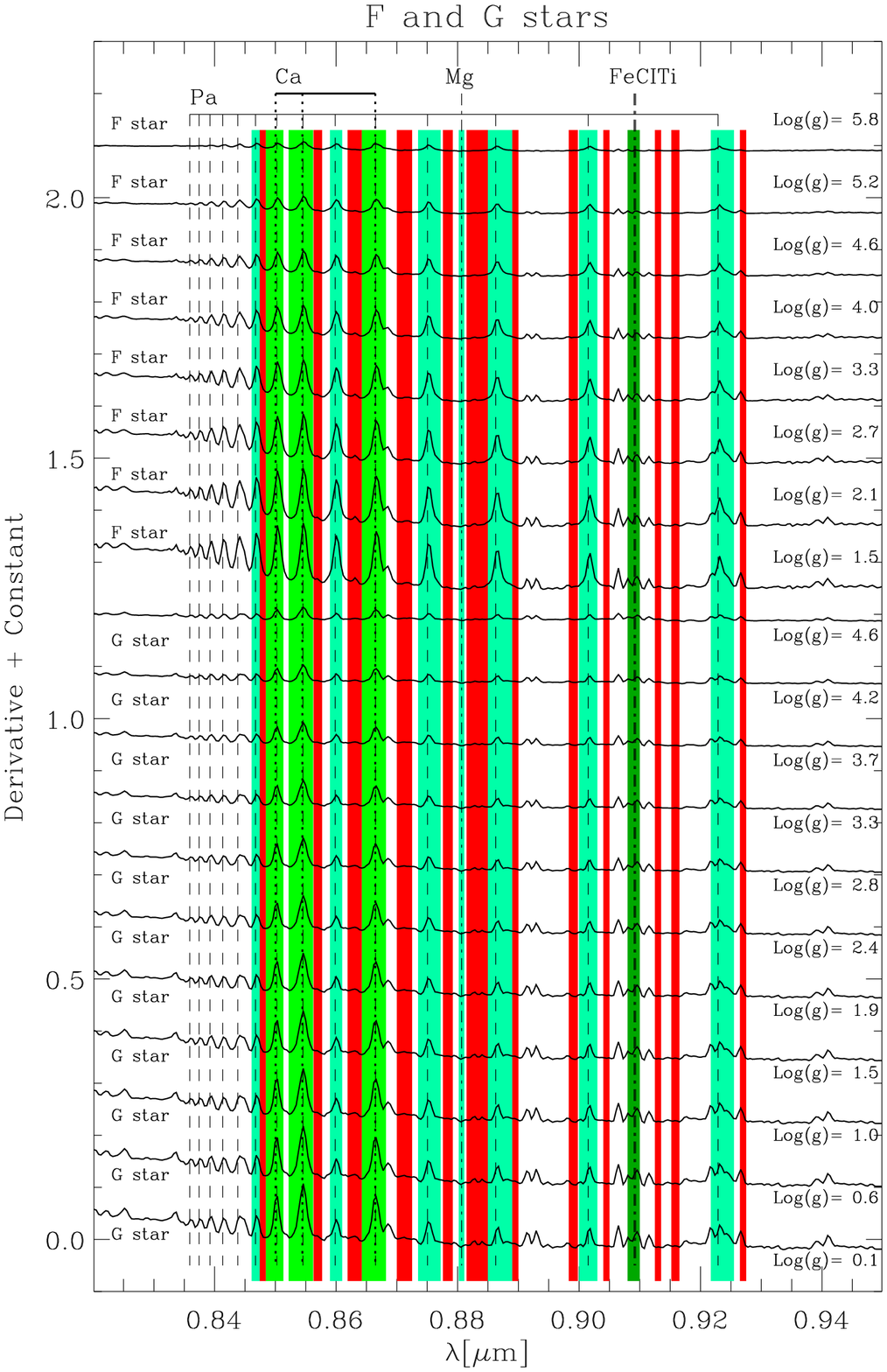}
\caption{$I$-band {\it sensitivity map} for surface gravity of F
(top) and G-type stars (bottom). The sensitivity map for
different gravity values is offset for displaying purposes and the
central values of the corresponding $\log\,(g)$ bins are
given. Symbols are as in Fig.\,\ref{fig:SGiant_fitted}.}
\label{fig:FG_Logg}
\end{figure*}

\begin{figure*}
\includegraphics[width=16truecm,angle=0]{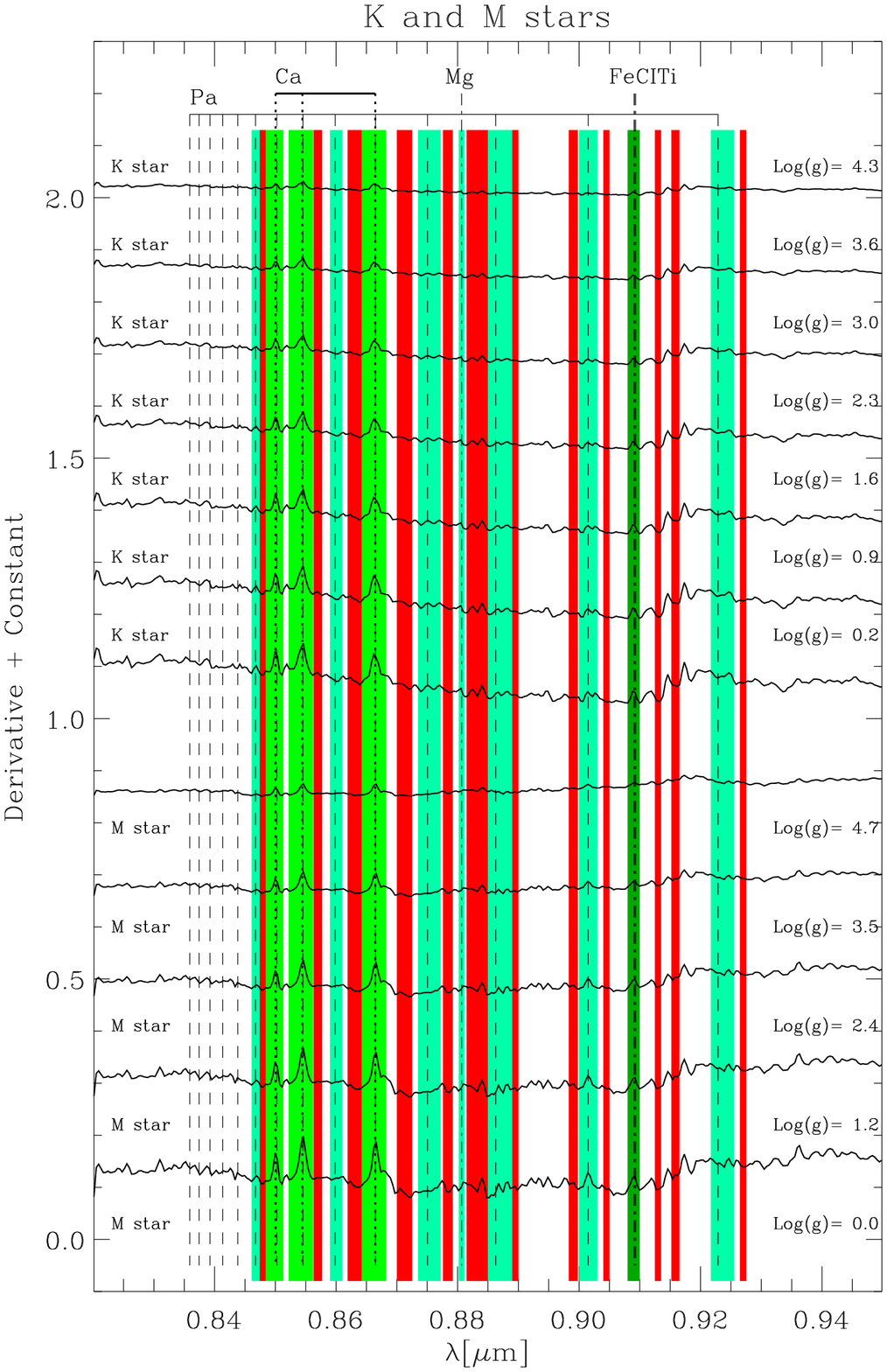}
\caption{As in Fig.\,\ref{fig:FG_Logg} but for K (top) and M-type
stars (bottom).}
\label{fig:KM_Logg}
\end{figure*}

The $I$-band sensitivity map of the surface gravity is displayed in
Figs.\,\ref{fig:FG_Logg} and \ref{fig:KM_Logg} for the F and G stars
and the K and M stars, respectively. For the purpose of this analysis
the sample was divided according to SpT and the L stars were excluded
because of insufficient gravity coverage. As expected, the CaT band
shows the strongest variation across all the SpTs, with positive peaks
at the core of the lines, consistent with CaT becoming weaker moving
from supergiant and giant towards dwarf stars. The sensitivity to
surface gravity (i.e., the strongest peaks) decreases for stars with
lowest $\log\,(g)$. The Paschen lines follow a similar trend but they
are useful only in F and, to some degree, in G stars, disappearing in
K and M stars, as discussed in the previous section. The Mg line
and FeClTi band show no noticeable variation. The change in the
overall shape of the sensitivity map in K and M stars for low
$\log\,(g)$ stars is probably due to the broad molecular features.

\subsection{Definition of new $I$-band indices}
\label{sec:ind_def_I}

The sensitivity map helped us to identify 11 $I$-band features
that can be used to measure the SpT and $\log\,(g)$. We defined
spectral indices in order to quantitatively analyze the relation
between the EW of these features and stellar physical parameters. The
indices consist of a central bandpass covering the feature of
interest, and two other adjacent bandpasses, at the red and blue
sides, tracing the local continuum. The central bandpass was selected
to include the peak of the sensitivity map and the continuum
bandpasses were placed on spectral regions where the sensitivity map
is (nearly) constant. For a perfectly uniform stellar library the
central bandpass can be defined to guarantee optimal extraction, but
this was not possible in the NIR region so we were conservative and
defined slightly wider bandpasses to assure that the spectral features
were fully encompassed. If possible, the combined width of the
continuum bandpasses was equal or larger than the width of the central
bandpass to avoid $S/N$ being degraded. For the Ca{~\sc ii} feature
the central bandpass overlaps with the Cen01 definition, whereas the
continuum bandpasses are different.  Six indices were defined to
characterize the Paschen series: three have narrower bandpasses than
in Cen01 and three have new bandpasses. The Mg feature has the
central bandpass overlapping with the \citet{Ray09} definition,
whereas the continuum bandpasses are different. Finally, we defined
a new index centered at $0.9090\,\mu$m to measure the combined
contribution of Fe{~\sc i}, Cl{~\sc i}, and Ti{~\sc i}. The bandpasses
of the $I$-band indices are listed in Table\,\ref{tab:index_def_I}.

\begin{table}[t]
\caption{Definition of the bandpasses of the $I$-band indices.}
\label{tab:index_def_I} 
\begin{tabular}{@{}l@{ }c@{ }c@{ }c@{}}
\hline
\hline
Index   & Element                          & Central bandpass   & Continuum bandpasses\\
        &                                  & ($\mu$m)         & ($\mu$m) \\
\hline
Pa1     & H{~\sc i} (n=3)                  &~0.8461--0.8474~&~0.8474--0.8484,~  0.8563--0.8577 \\
Ca1     & Ca{~\sc ii}                      &~0.8484--0.8513~&~0.8474--0.8484,~  0.8563--0.8577 \\
Ca2     & Ca{~\sc ii}                      &~0.8522--0.8562~&~0.8474--0.8484,~  0.8563--0.8577 \\
Pa2     & H{~\sc i} (n=3)                  &~0.8577--0.8619~&~0.8563--0.8577,~  0.8619--0.8642 \\
Ca3     & Ca{~\sc ii}                      &~0.8642--0.8682~&~0.8619--0.8642,~  0.8700--0.8725 \\
Pa3     & H{~\sc i} (n=3)                  &~0.8730--0.8772~&~0.8700--0.8725,~  0.8776--0.8792 \\
Mg      & Mg{~\sc i}                       &~0.8802--0.8811~&~0.8776--0.8792,~  0.8815--0.8850 \\
Pa4     & H{~\sc i} (n=3)                  &~0.8850--0.8890~&~0.8815--0.8850,~  0.8890--0.8900 \\
Pa5     & H{~\sc i} (n=3)                  &~0.9000--0.9030~&~0.8983--0.8998,~  0.9040--0.9050 \\
FeClTi~~&~Fe{~\sc i},Cl{~\sc i},Ti{~\sc i}~&~0.9080--0.9100~&~0.9040--0.9050,~  0.9125--0.9135 \\
Pa6     & H{~\sc i} (n=3)                  &~0.9217--0.9255~&~0.9152--0.9165,~  0.9265--0.9275 \\
\hline                 
\end{tabular}
\end{table}

The EW is defined as
\begin{equation} 
EW = \int_{\lambda1}^{\lambda2} (1 - F_{{\rm line}} / F_{{\rm cont}}) \  d\lambda,
\end{equation} 
where $F_{{\rm line}}$ is the flux density of the observed spectrum
$F(\lambda)$ inside the line bandpass between $\lambda_1$ and
$\lambda_2$, $F_{\rm {cont}}$ is the value of the local continuum at
the central wavelength of the line bandpass as obtained by linearly
interpolating between the two continuum bandpasses, and
$\Delta\lambda=\lambda_2-\lambda_1$ is the width of the line bandpass.
The measurements were performed under the IDL environment using a
specially developed script. The EW is measured directly on the
observed spectrum. To derive the EW errors we first computed the rms
of the residual of the spectrum and the continuum (where the continuum
is assumed to be the straight line interpolating the two continuum
bandpasses) in the two continuum regions.  We then estimated the EW
errors by means of Monte Carlo simulations on the observed spectrum
taking as noise the root mean square (rms) above derived. We also
measured the CaT, PaT and CaT* indices as defined by Cen01. The
results are listed in Tables\,\ref{tab:index_mis_I_1} and 
\ref{tab:index_mis_I_2}


\section{Spectral diagnostics in the $I$ band}
\label{sec:SpT_diag_I}

\subsection{Spectral diagnostics for the spectral type}

\begin{figure*}
\includegraphics[width=9truecm,height=8truecm,angle=0]{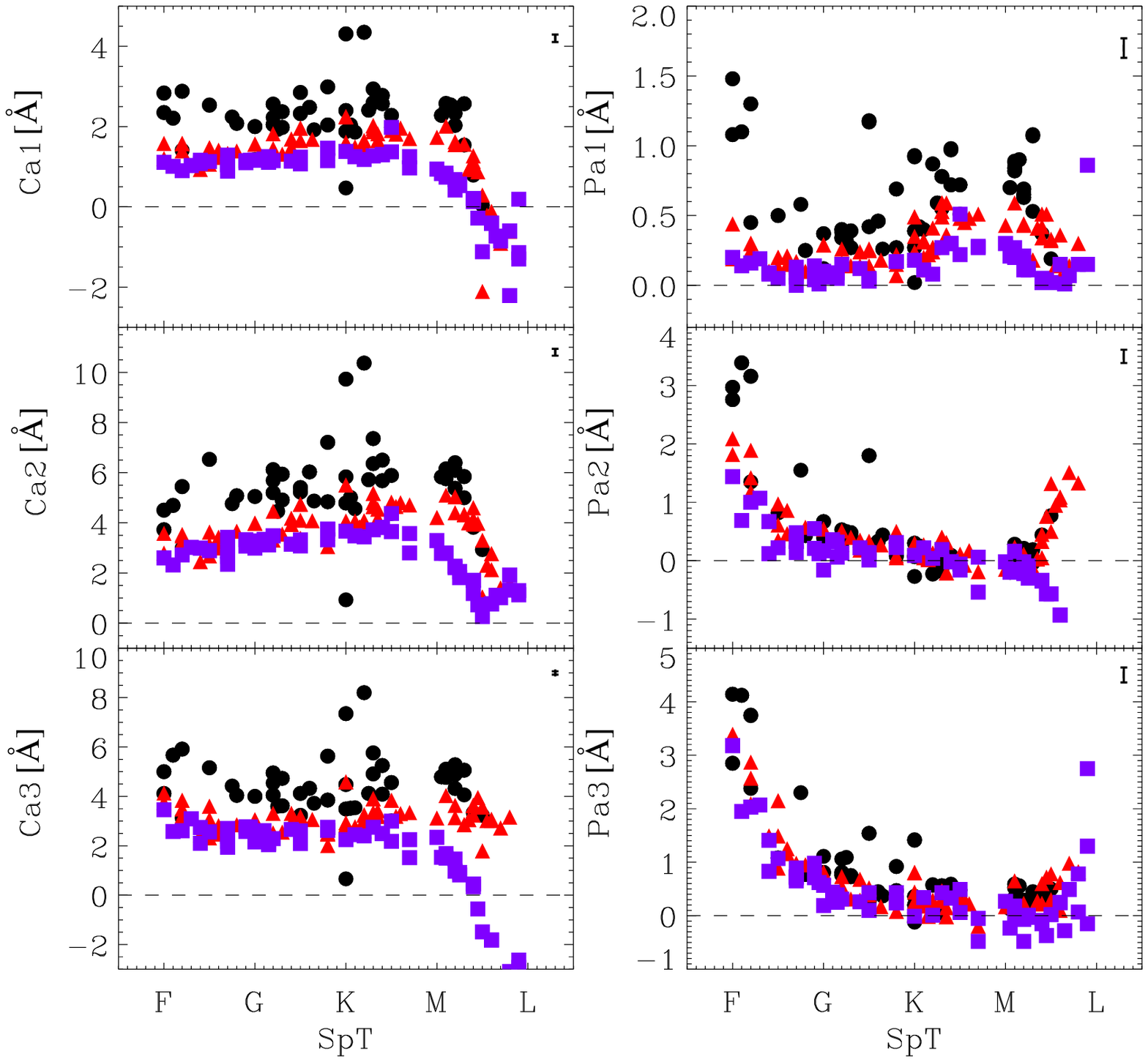}
\includegraphics[width=9truecm,height=8truecm,angle=0]{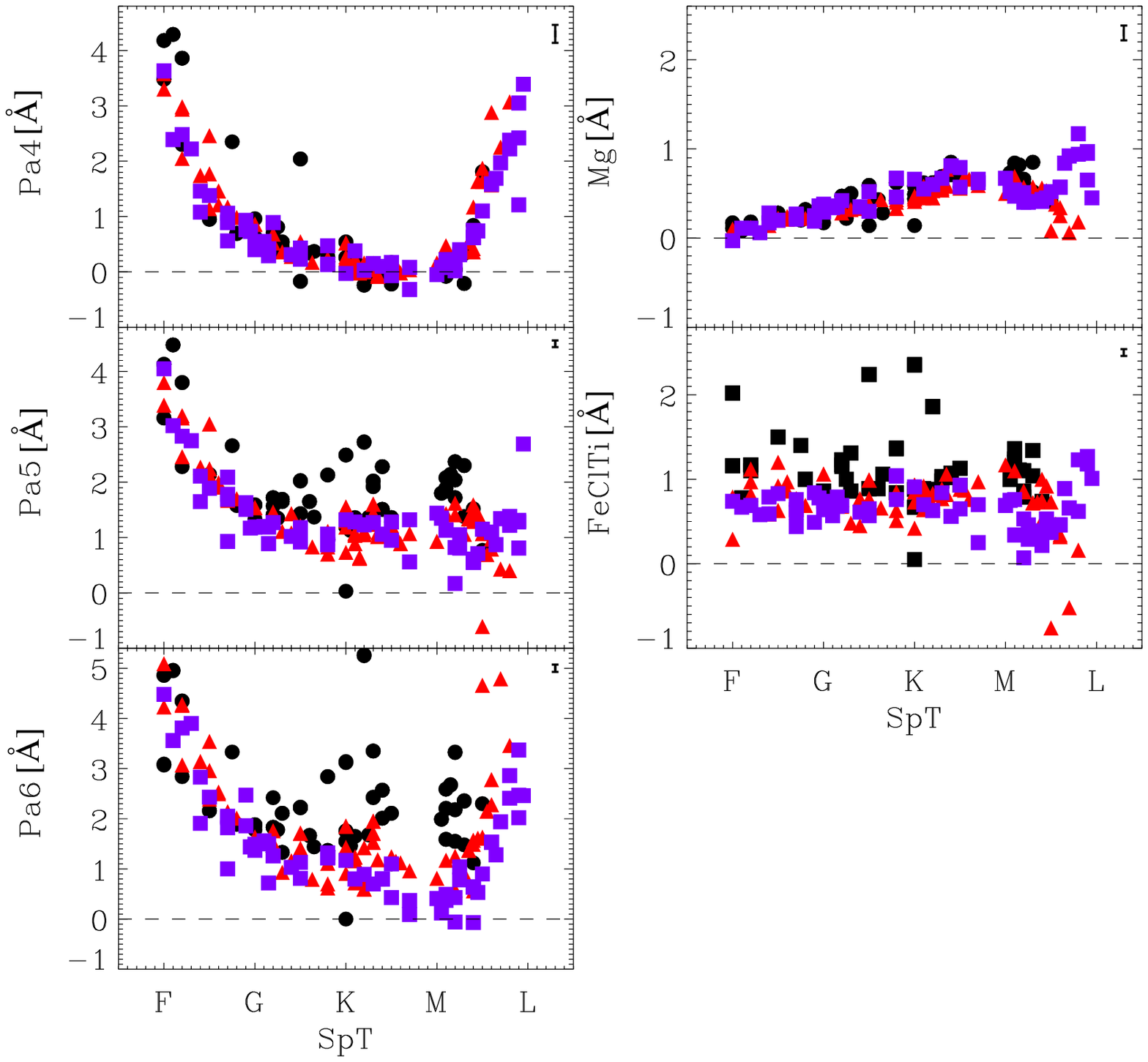}\\
\includegraphics[width=9truecm,height=8truecm,angle=0]{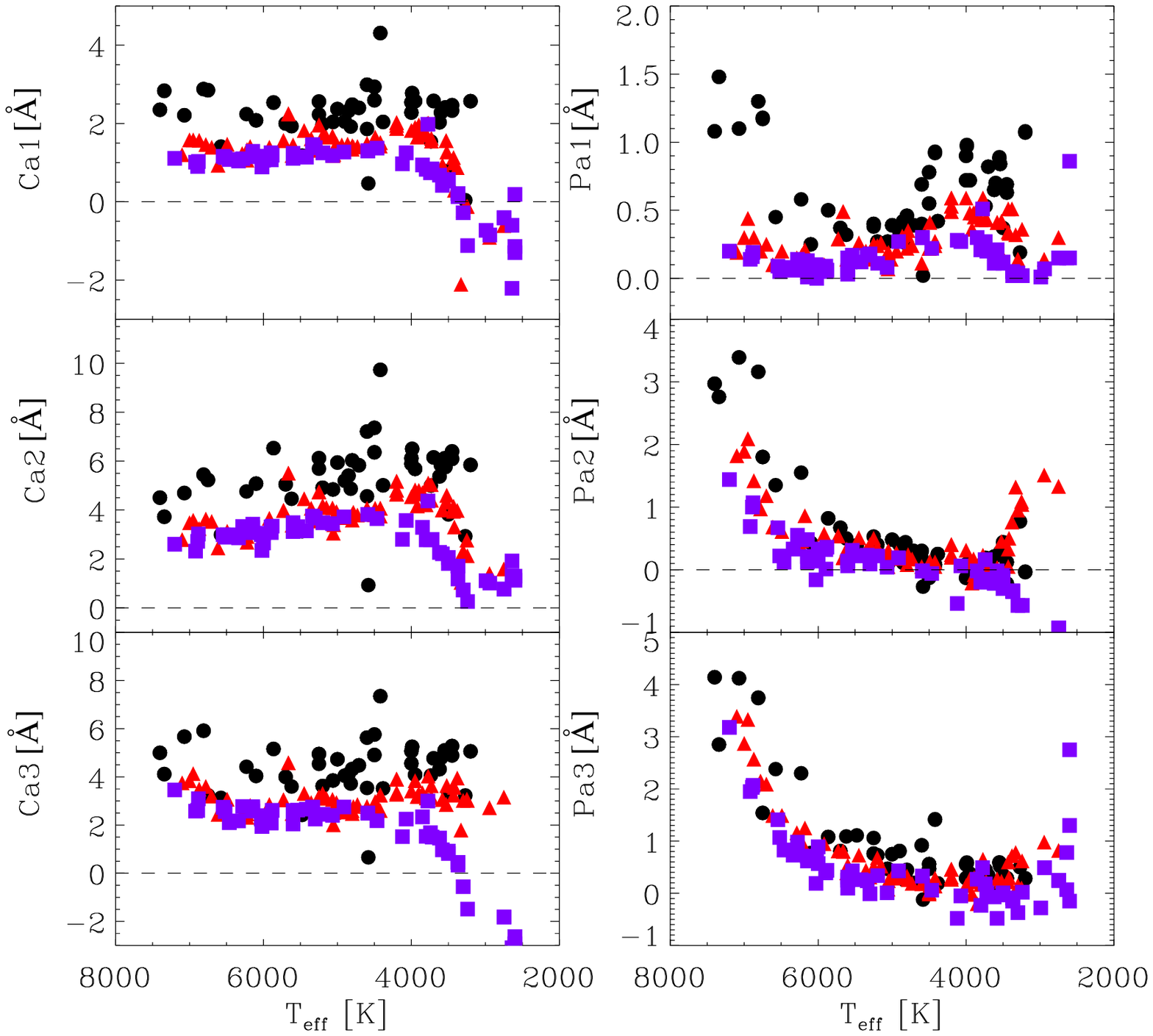}
\includegraphics[width=9truecm,height=8truecm,angle=0]{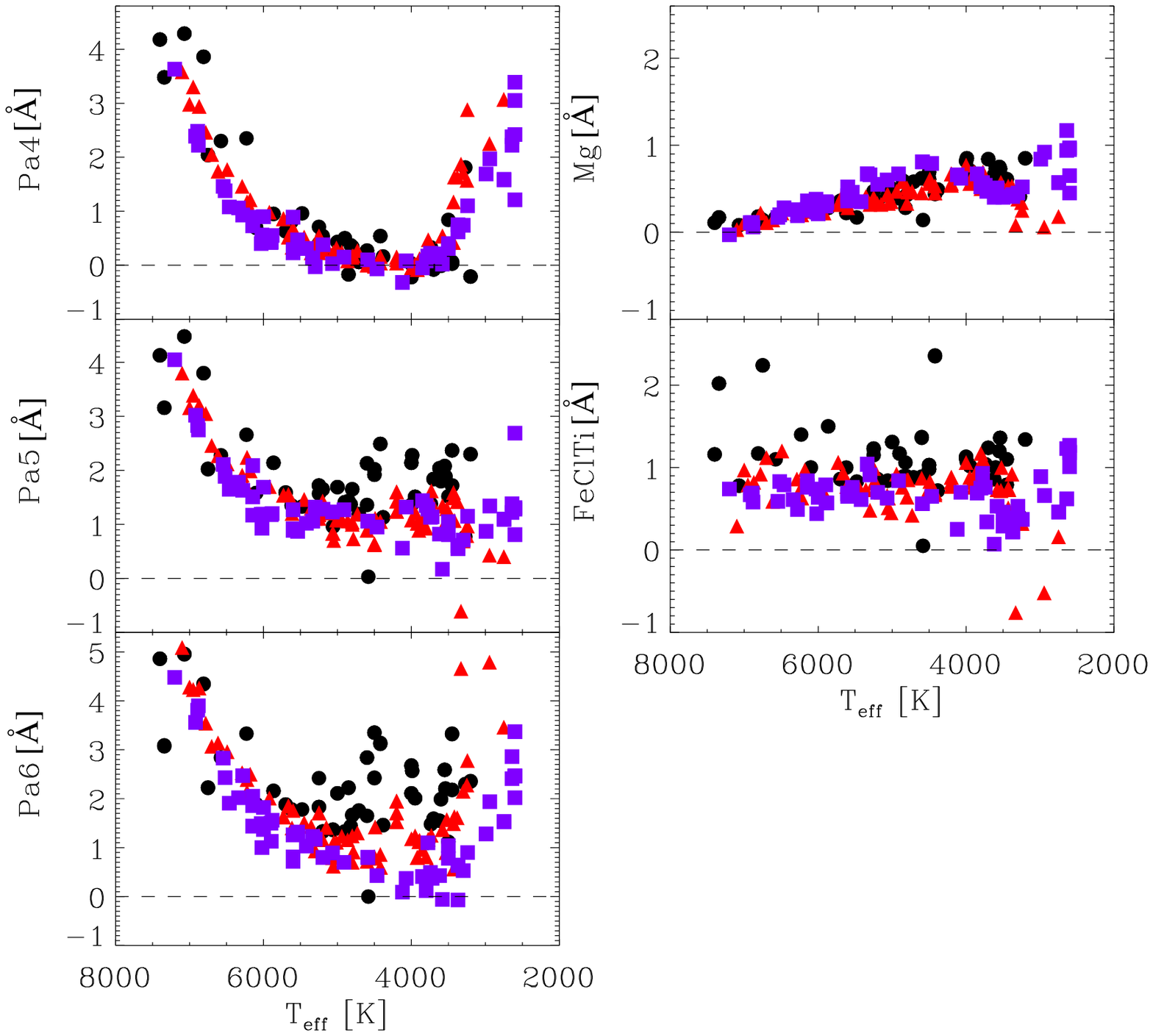}
\caption{Equivalent width of the $I$-band indices as function of
  spectral type (top panels) and effective temperature (bottom
  panels). The different symbols correspond to the supergiant
  (circles), giant (triangles), and dwarf stars (squares),
  respectively. For each index on the right side of the plot the median
  of the errors is shown (top panels).}
\label{fig:Ind_Spt}
\end{figure*}

The new $I$-band indices are plotted as a function of the SpT and
$T_{\rm eff}$ in Fig.\,\ref{fig:Ind_Spt}. The plots confirm that the
CaT lines are not sensitive to the SpT for stars from F to early-M
type, (although they have some sensitivity to the luminosity class,
i.e., $\log\,(g)$, see Sect.\,\ref{Sec:SpDiag}) and show a marked
decrease for SpT later than M, vanishing beyond the M5 type. The
negative values are due to the broad TiO bands affecting the
continuum, particularly in the dwarf and, to some extent, in giant
stars.

Pa1 shows negligible variation with SpT, whereas Pa2 and the other
H{~\sc i} lines decrease, most notably from F to early-G stars.  The
sharp rise beyond the M type is due to molecular contamination, which
is strongest in Pa4. The scatter is the largest for the reddest lines,
probably due to the increasing of the sky background and worsening of
the atmospheric transmission. Mg shows a constant increasing with
SpT. Finally, the FeClTi band appears insensitive to the SpT for all
luminosity classes.

We conclude that the sensitivity map method correctly recovers the
different behavior of Mg, CaT and Paschen lines and it allows to
determine the SpT of stars. However, we notice that the CaT in
supergiant stars is characterized by a larger scatter than in dwarf
and giant stars. This is not due to a metallicity effect (see
Sect.\,\ref{Sec:SpDiag}).

\subsection{Spectral diagnostics for the surface gravity and
    metallicity}
\label{Sec:SpDiag}

The Paschen indices show no trend with surface gravity. The
sensitivity map indicates a mild decrease of the Paschen lines with
increasing $\log\,(g)$ for F and G stars, but the large EW scatter for
the hotter sample stars prevented us from measuring any gradient. K
and M stars are characterized by a smaller EW scatter and we found
constant Paschen values as expected. The Mg and FeClTi feature
do not show any correlation with surface gravity as expected from
the sensitivity map.

Weak trends are observed for Ca1 and FeClTi but the narrow metallicity
range of the stellar library prevented us from drawing firm conclusions 
whether these indices can be used to derive [Fe/H]. The CaT index 
is indeed a well understood metallicity indicator \citep{Ter89,Tol09},
although it degenerates at higher metallicities \citep{Vaz03}.

\begin{figure*}
\includegraphics[width=9truecm,height=8truecm,angle=0]{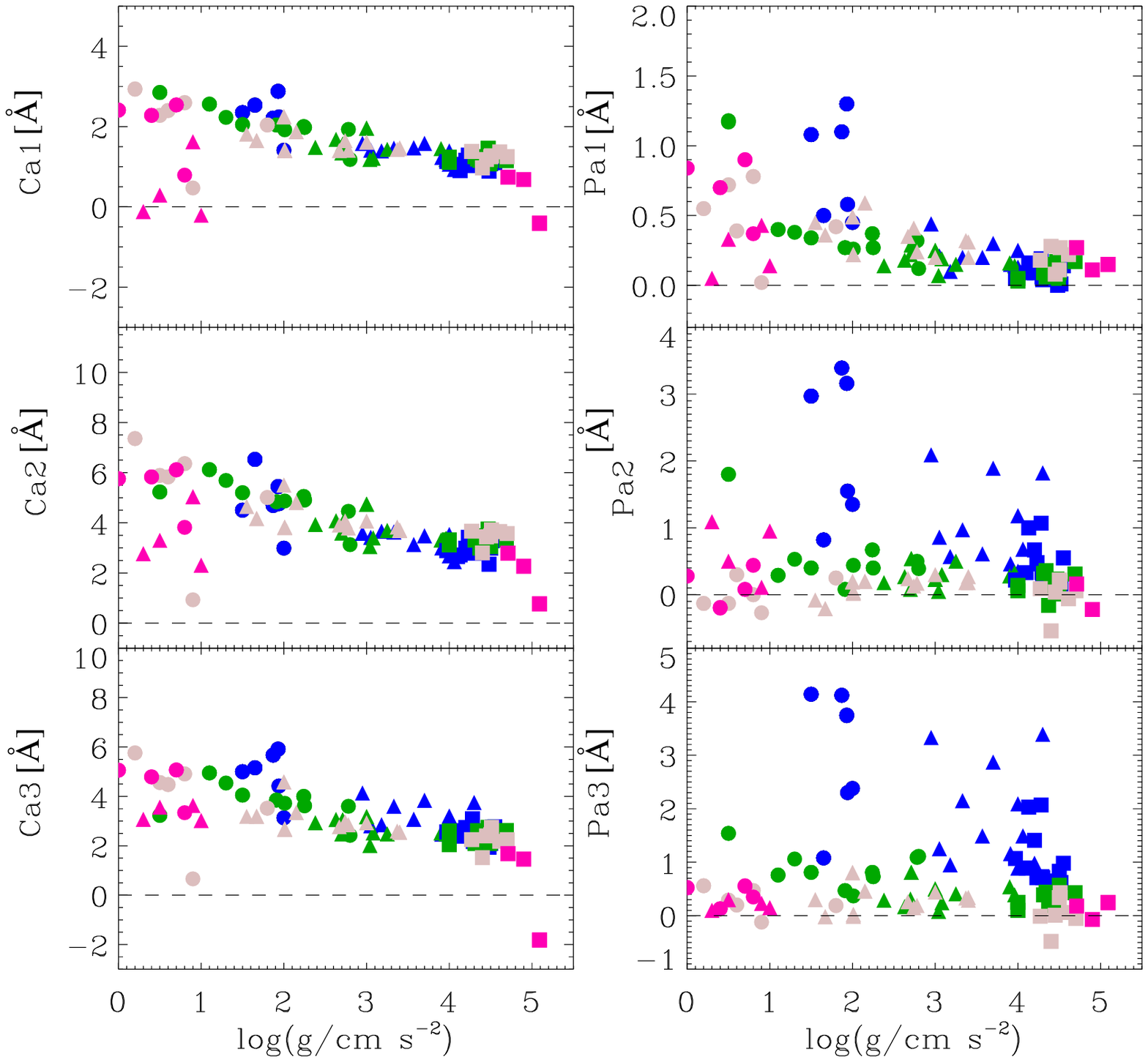}
\includegraphics[width=9truecm,height=8truecm,angle=0]{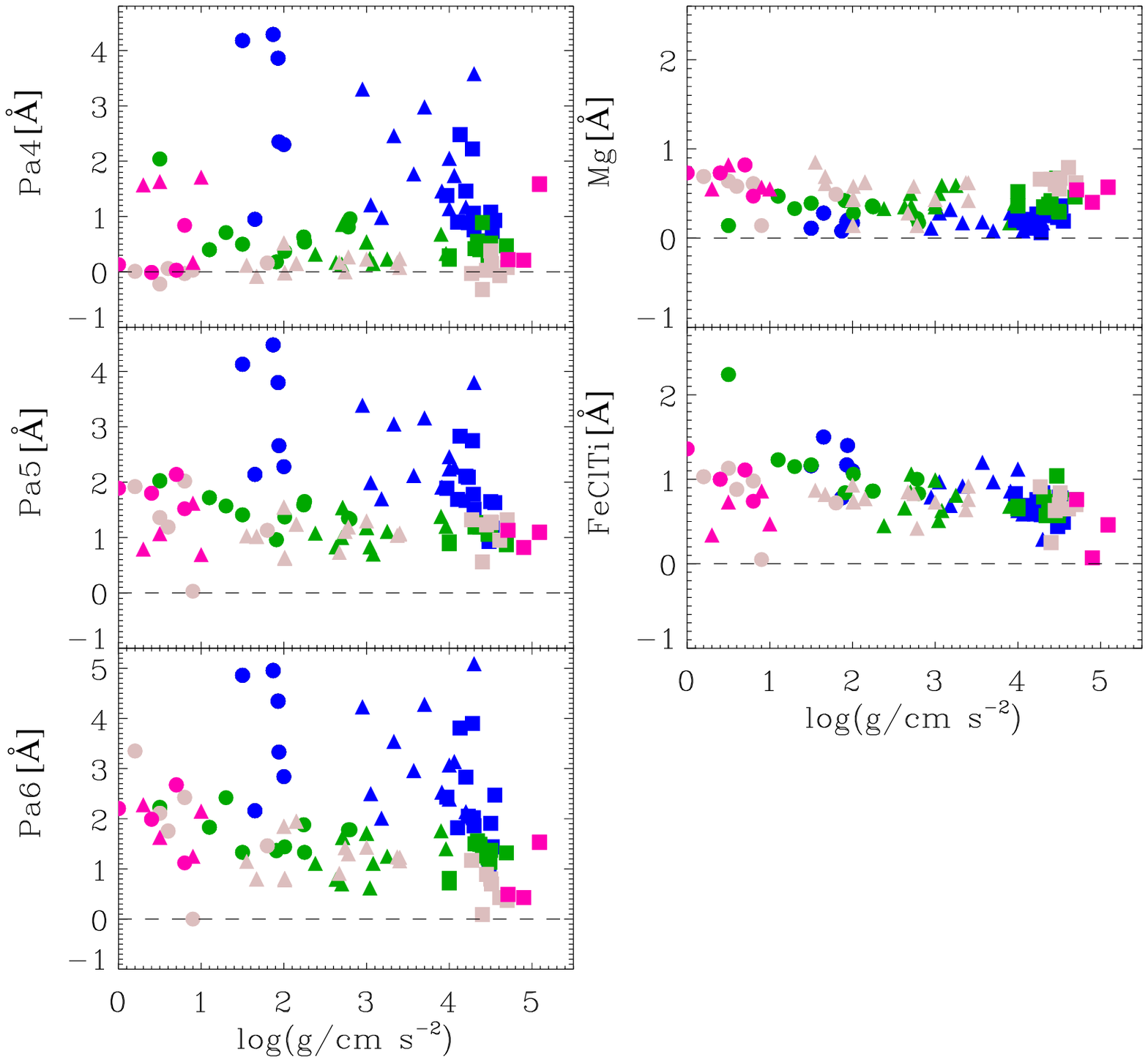}\\
\includegraphics[width=9truecm,height=8truecm,angle=0]{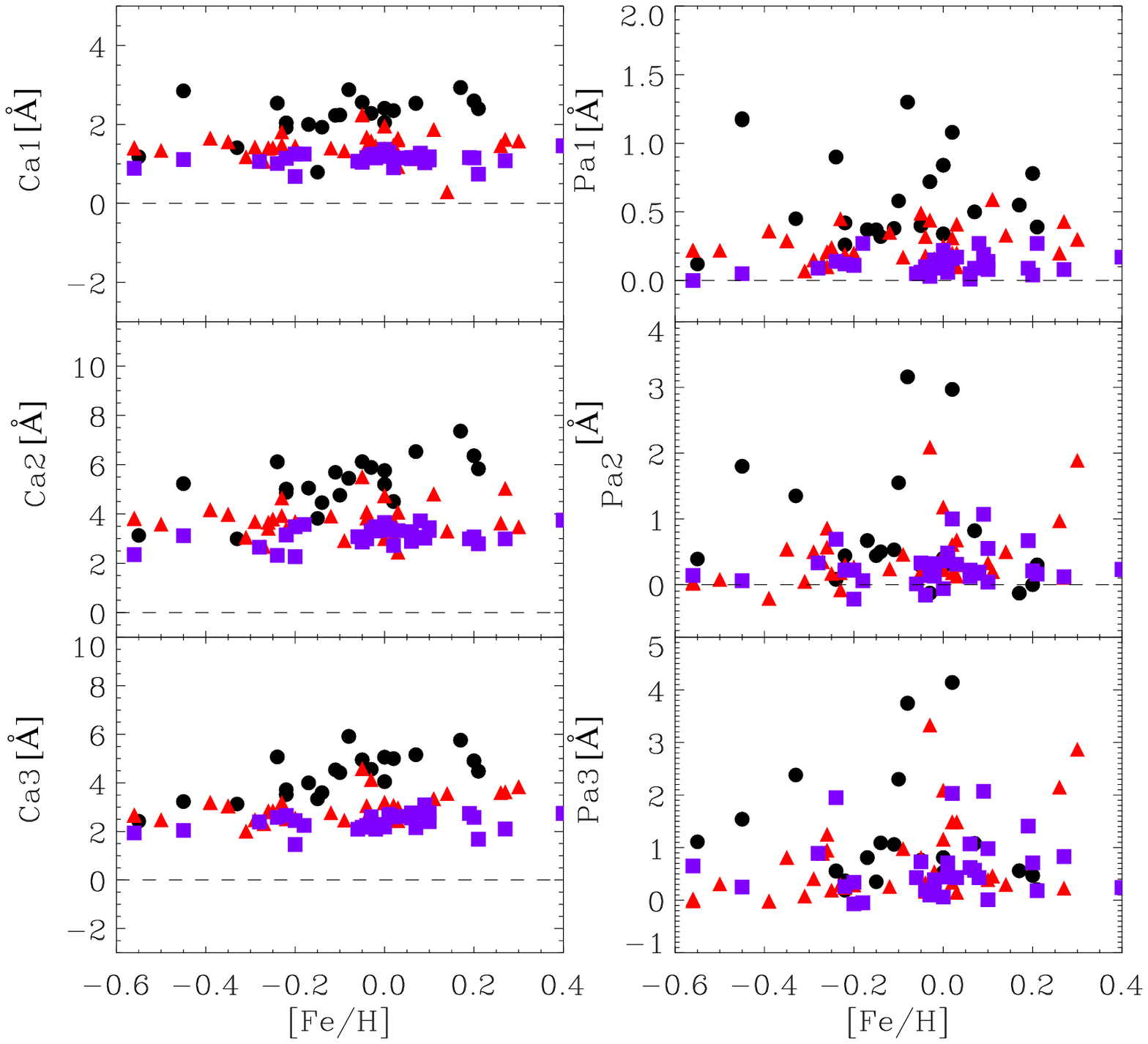}
\includegraphics[width=9truecm,height=8truecm,angle=0]{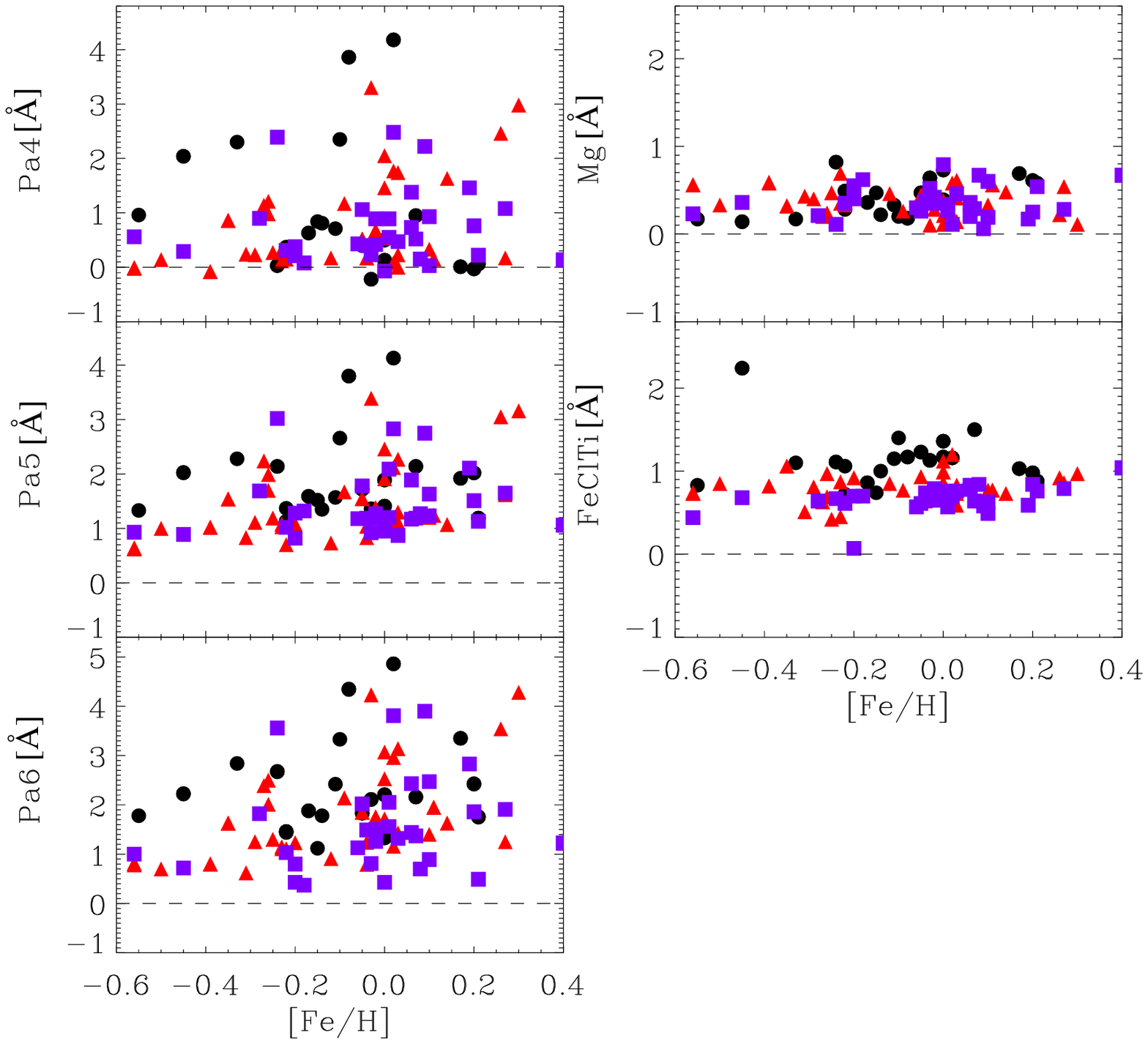}
\caption{Equivalent width of the $I$-band indices as function of
  surface gravity (top panels) and metallicity (bottom
  panels). Symbols for supergiant, giant, and dwarf stars and errors
  are as in Fig.\,\ref{fig:Ind_Spt}. Different colors in top panels
  correspond to F (blue), G (green), K (grey), and M stars (pink),
  respectively.}
\label{fig:Ind_Gr}
\end{figure*}

\clearpage

\section{Spectral indices in the $K$ band}
\label{sec:sp_ind_K}

\subsection{Main $K$-band spectral features}

The same analysis described in the previous sections for the $I$ band
was carried out for the $K$ band ($1.92-2.40\,\mu$m), which also
contains some well-studied features. The most prominent are the series
of Ca{~\sc i} lines at 1.95$\,\mu$m, Na{~\sc i} doublet at
2.21$\,\mu$m, Ca{~\sc i} doublet at 2.26$\,\mu$m, and the series of
first-overtone bandheads of $^{12}$CO extending redward of
2.29$\,\mu$m and of $^{13}$CO extending redward of 2.34$\,\mu$m
\citep{Cus05}. The F stars show H{~\sc i} absorption lines of the
Bracket series. They are the Br$\delta$ at 1.94$\,\mu$m, which falls
in a wavelength region of moderate telluric absorption and Br$\gamma$
at 2.16$\,\mu$m. The metal features and the CO are known to increase
with SpT, whereas the H{~\sc i} lines decrease with SpT
\citep[e.g.,][]{Iva00}. Finally, broad water-absorption bands appear
on both sides of the $K$ band in late-M type stars and they smoothly
decrease in strength from supergiants through giants to dwarfs
\citep[e.g.,][]{Lan07}.

\subsection{Sensitivity map for the spectral type}
\label{sec:sens_fun1_K}

\begin{figure*}
\includegraphics[width=16truecm,angle=0]{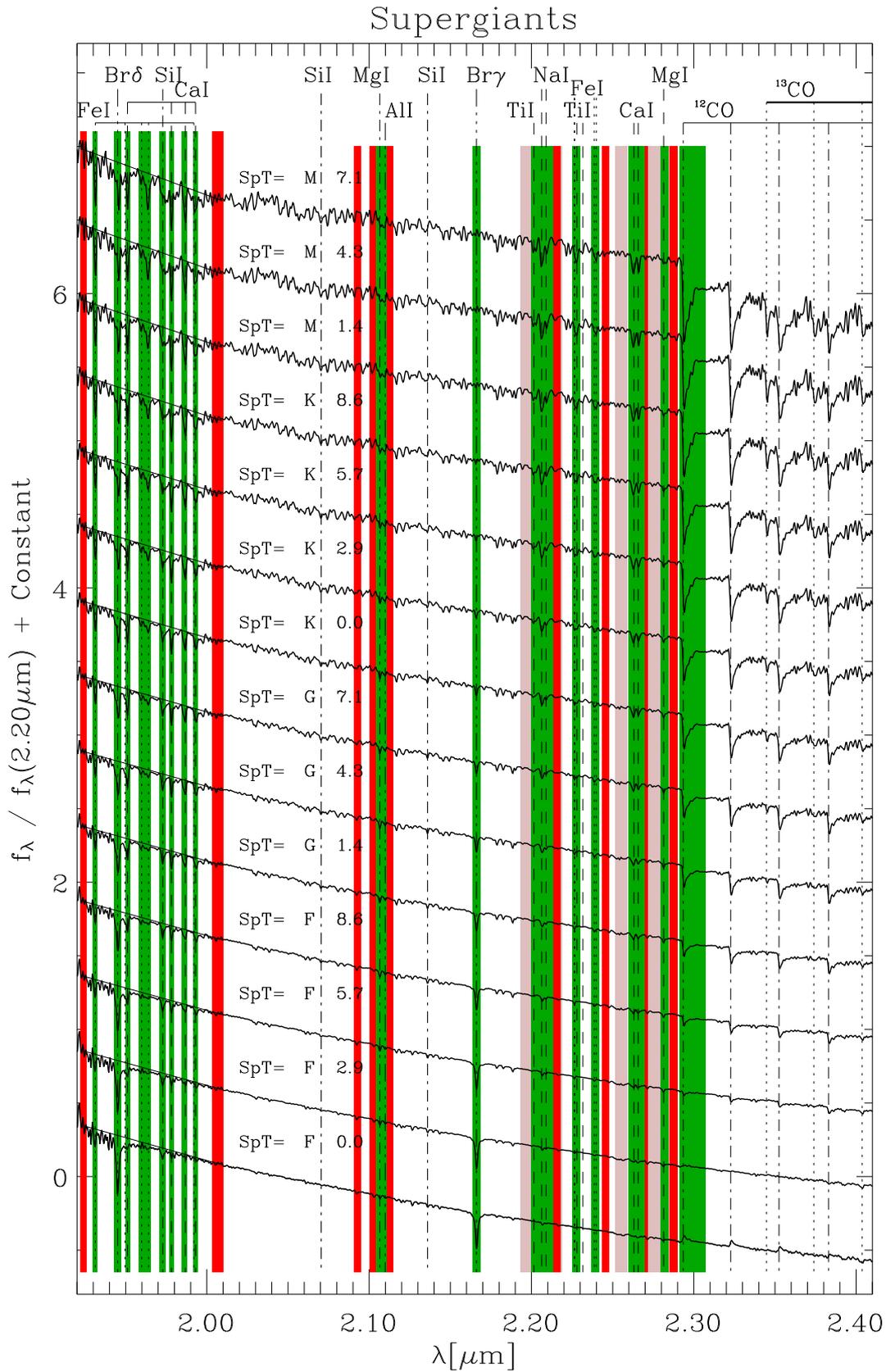}
\caption{$I$-band {\it model spectrum} of supergiant stars obtained by
 fitting at each wavelength the flux-normalized (at 2.20$\,\mu$m)
 sample spectra along SpTs. The model spectrum for different SpTs is
 offset for displaying purposes and the SpT is given. The green
 regions mark the bandpasses of the newly defined indices and, the red
 and grey regions, mark their adjacent continuum as defined in this
 paper and in literature respectively (see
 Table\,\ref{tab:index_def_K}). Some relevant absorption features are
 marked.}
\label{fig:SGiant_fitted_K}
\end{figure*}

\begin{figure*}
\includegraphics[width=16truecm,angle=0]{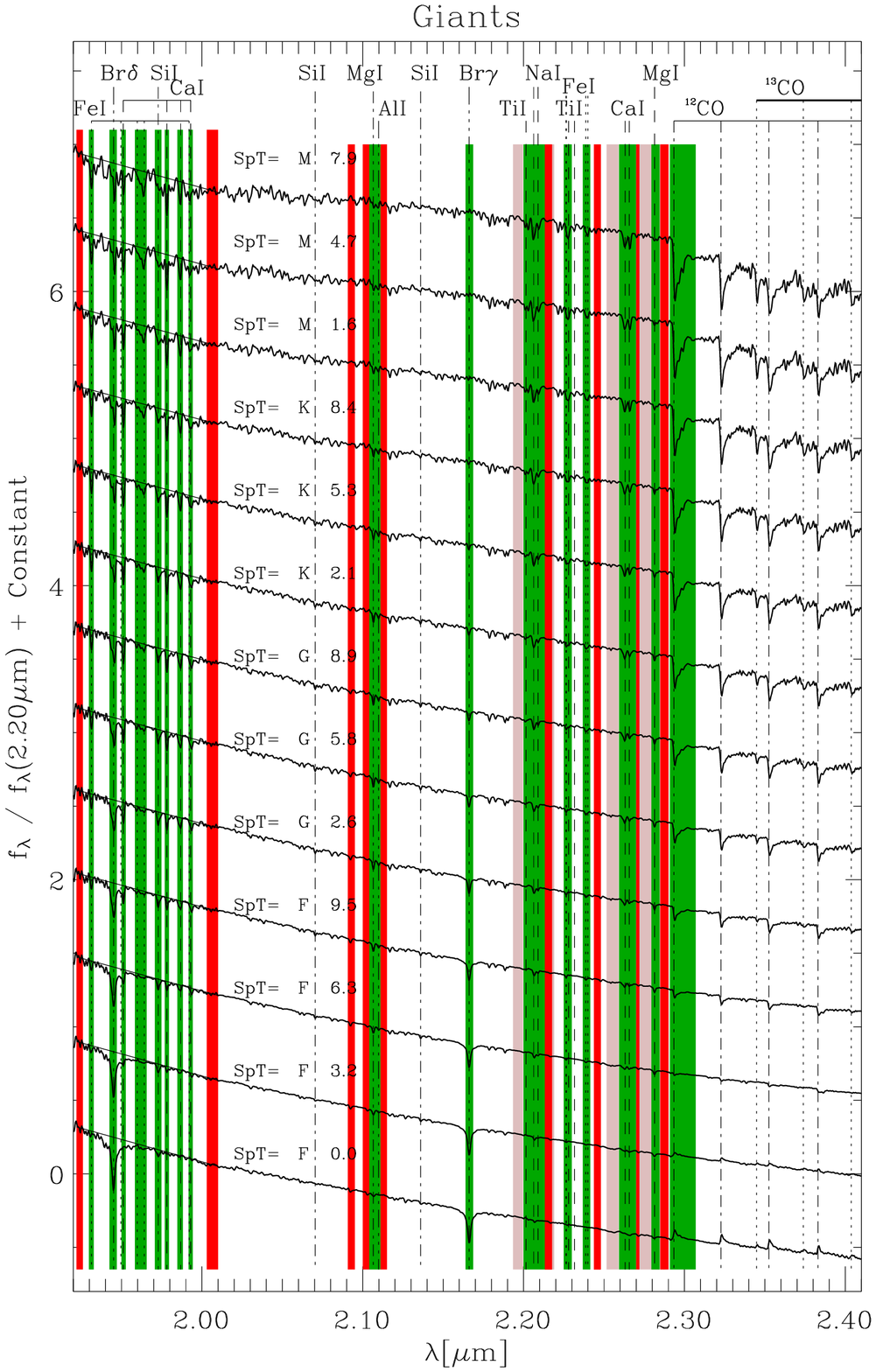}
\caption{As in Fig.\,\ref{fig:SGiant_fitted_K} but for giant stars.}
\label{fig:Giant_fitted_K}
\end{figure*}

\begin{figure*}
\includegraphics[width=16truecm,angle=0]{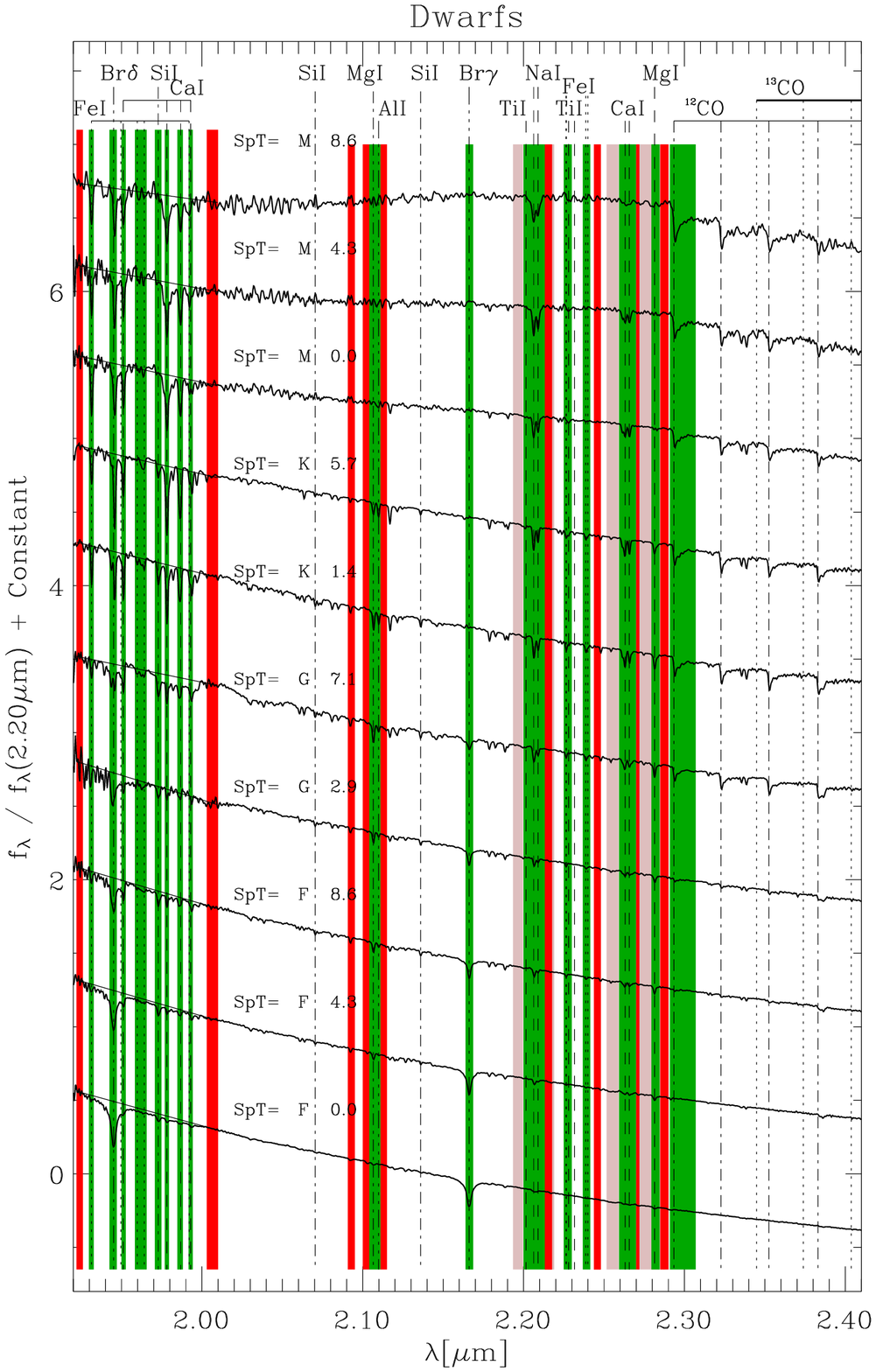}
\caption{As in Fig.\,\ref{fig:SGiant_fitted_K} but for dwarf stars.}
\label{fig:Dwarf_fitted_K}
\end{figure*}

\begin{figure*}
\includegraphics[width=16truecm,angle=0]{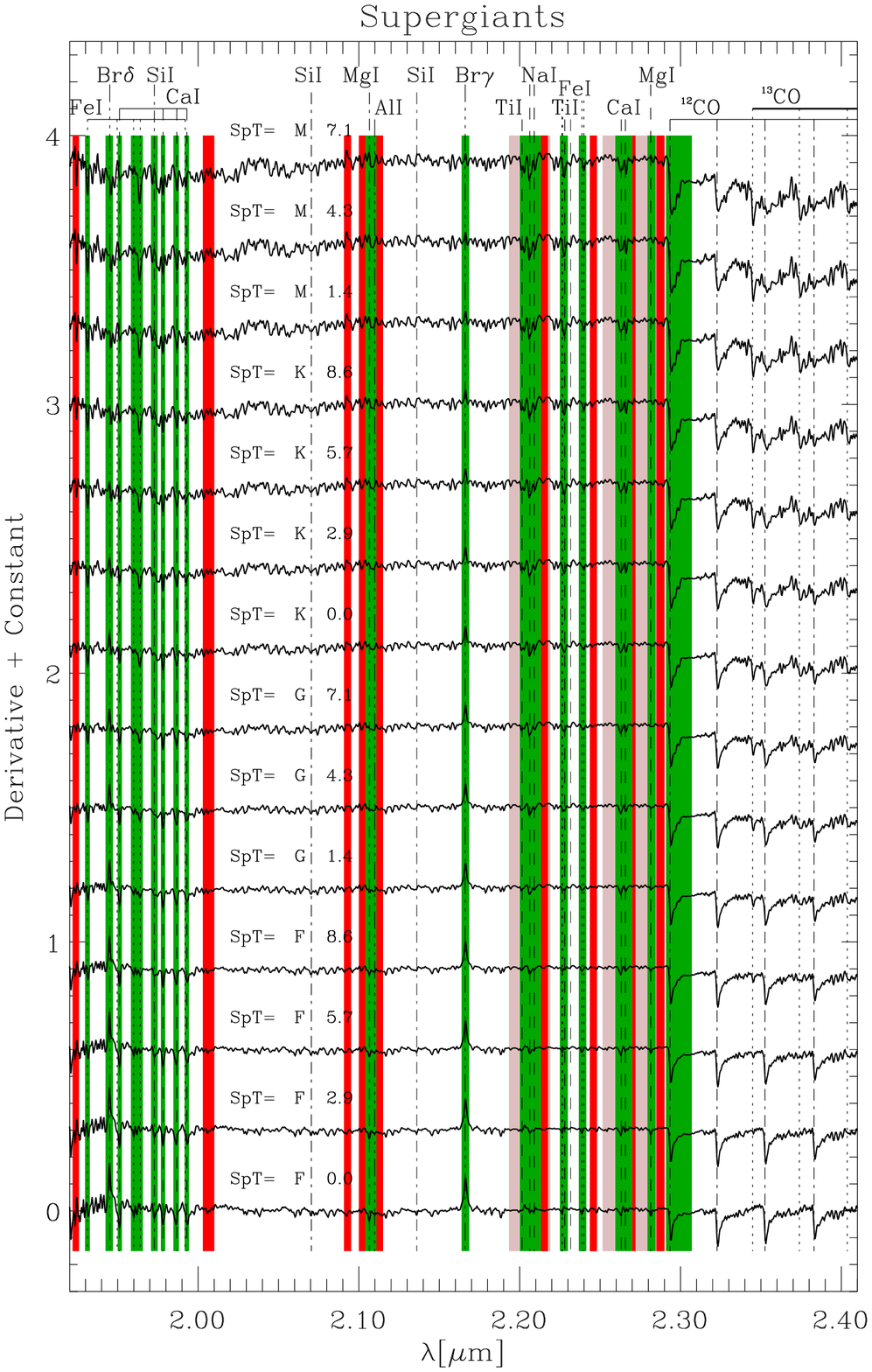}
\caption{$K$-band {\it sensitivity map} for SpT of supergiant
    stars.  The sensitivity map for different SpTs is offset for
    displaying purposes and the SpT is given. Symbols are as in
    Fig.\,\,\ref{fig:SGiant_fitted_K}}
\label{fig:SupGian_SpT_K}
\end{figure*}

\begin{figure*}
\includegraphics[width=16truecm,angle=0]{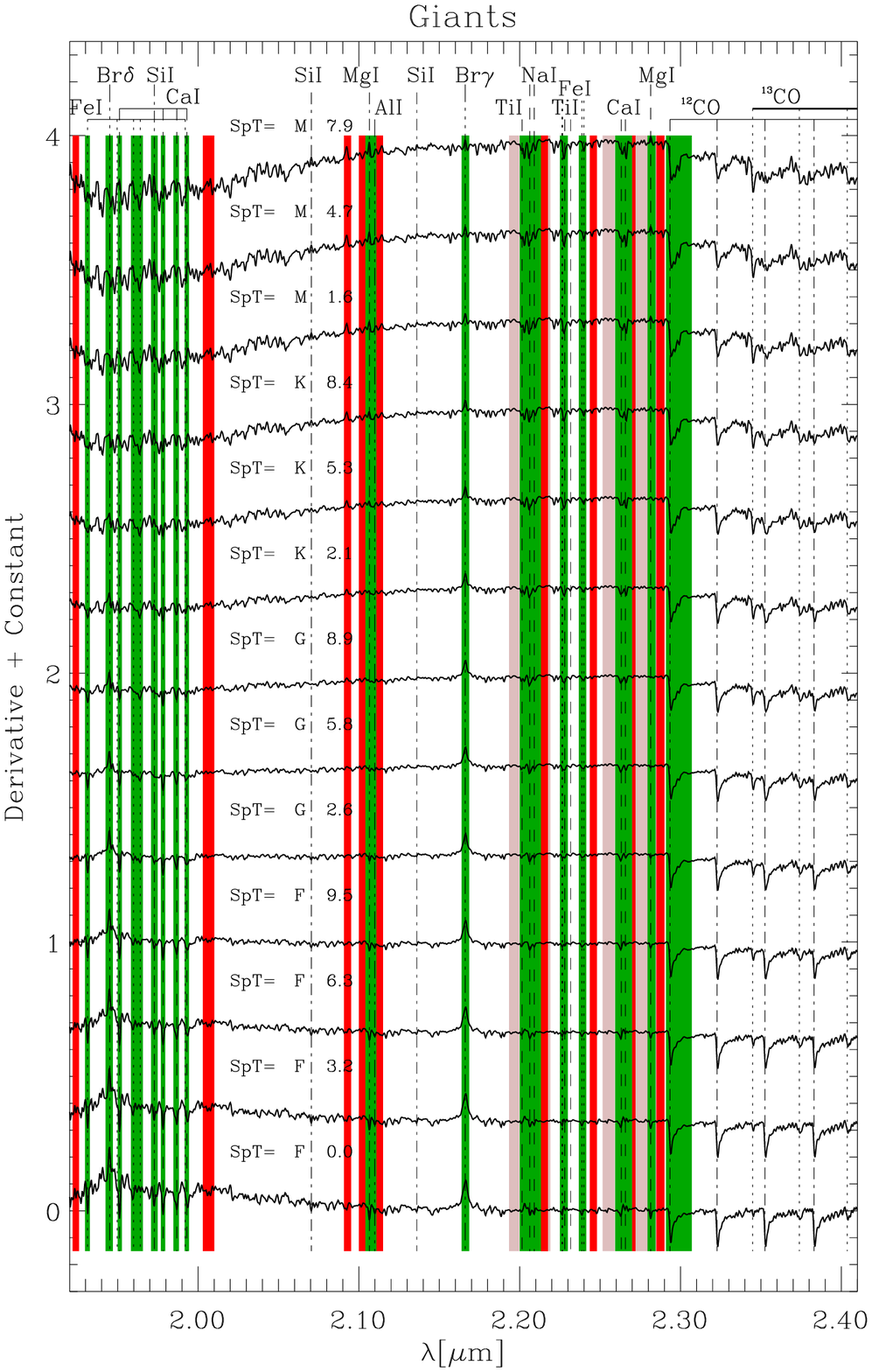}
\caption{As in Fig.\,\ref{fig:SupGian_SpT_K} but for giant stars.}
\label{fig:Gian_SpT_K}
\end{figure*}

\begin{figure*}
\includegraphics[width=16truecm,angle=0]{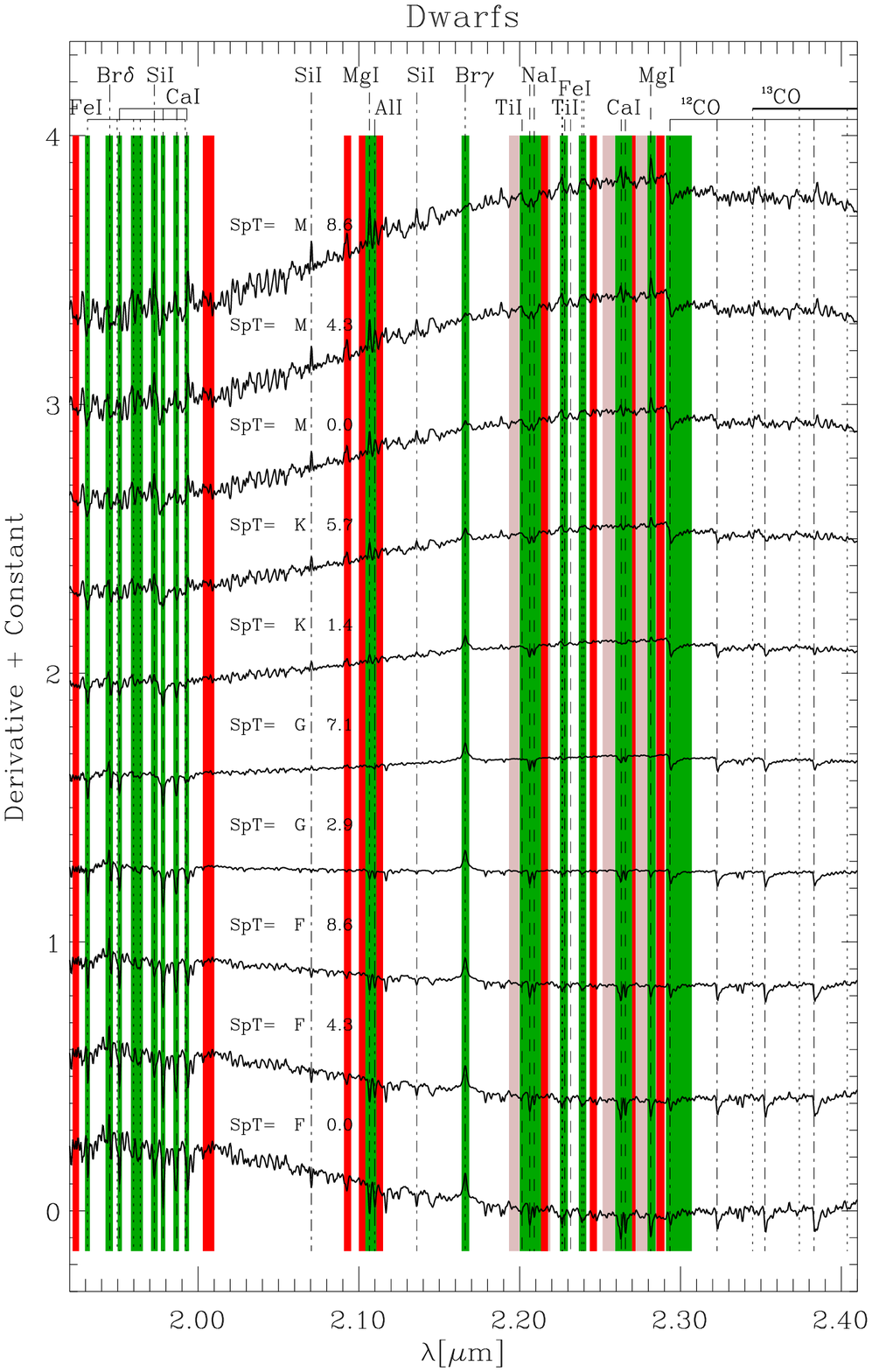}
\caption{As in Fig.\,\ref{fig:SupGian_SpT_K} but for dwarf stars.}
\label{fig:Dwarf_SpT_K}
\end{figure*}

The $K$-band model spectrum and sensitivity map as a function of SpT
are shown in Figs.\,\ref{fig:SGiant_fitted_K},
\ref{fig:Giant_fitted_K}, and \ref{fig:Dwarf_fitted_K} and in
Figs.\,\ref{fig:SupGian_SpT_K}, \ref{fig:Gian_SpT_K},
\ref{fig:Dwarf_SpT_K} for supergiant, giant, and dwarf stars,
respectively. Each luminosity class was considered separately. The
analysis of the supergiant stars can be used to summarize the features
present in the $K$ band and done for the $I$ band. The bluest part of
the $K$ band contains a complex of Fe{~\sc i} and Ca{~\sc i} lines at
about $1.95$--$1.99\,\mu$m. The sensitivity map shows dips with
variable strength, corresponding to an increase of the line strengths
from F0 to late-F stars, a plateau for the F--G types, and a further
increase for SpTs from K5 to M7.

The sensitivity map of the Si{~\sc i} features shows a dip for the F
stars and it is flat -- within the scatter -- for later SpTs,
indicating that the line looses its sensitivity to $T_{\rm eff}$ for
redder stars. The Br$\gamma$ has a peak which decreases with SpT 
from F0.0 to K8.6, because the flux in the line rises towards 
later types faster than the neighboring continuum, making the line 
weaker. Br$\delta$ also corresponds to a peak in the sensitivity map
and the line almost completely disappears after K2-5 type. The blend
of Mg{~\sc i} at 2.106$\,\mu$m and Al{~\sc i} at 2.110$\,\mu$m
exhibits a shallow dip in the sensitivity map, {\it i.e.} it
increases mildly with the SpT.  The EWs of the Na{~\sc i} doublet at
2.21$\,\mu$m and Ca{~\sc i} doublet at 2.26$\,\mu$m follow the pattern
of the previously discussed Ca lines. The two Fe{~\sc i} at 2.23 and
2.24$\,\mu$m are not sensitive to the SpT in the range between F and K
and they increase only for M stars. The Mg{~\sc i} at 2.28$\,\mu$m
shows no marked variations for different SpTs. The $^{12}$CO at
2.29$\,\mu$m shows a dip in the sensitivity map and it strongly
increases with the SpT.

Summarizing, the spectral features in the supergiant, giant, and dwarf
stars show similar behavior -- perhaps, with slightly different
strengths of the gradients. A difference may be noted in the overall
shape of the derivatives due to the broad water vapor absorption at
the blue and the red edges of the $K$ band in the late-K and M-type
dwarf stars.

\subsection{Sensitivity map for the surface gravity}
\label{sec:sens_fun2_K}

\begin{figure*}
\includegraphics[width=16truecm,angle=0]{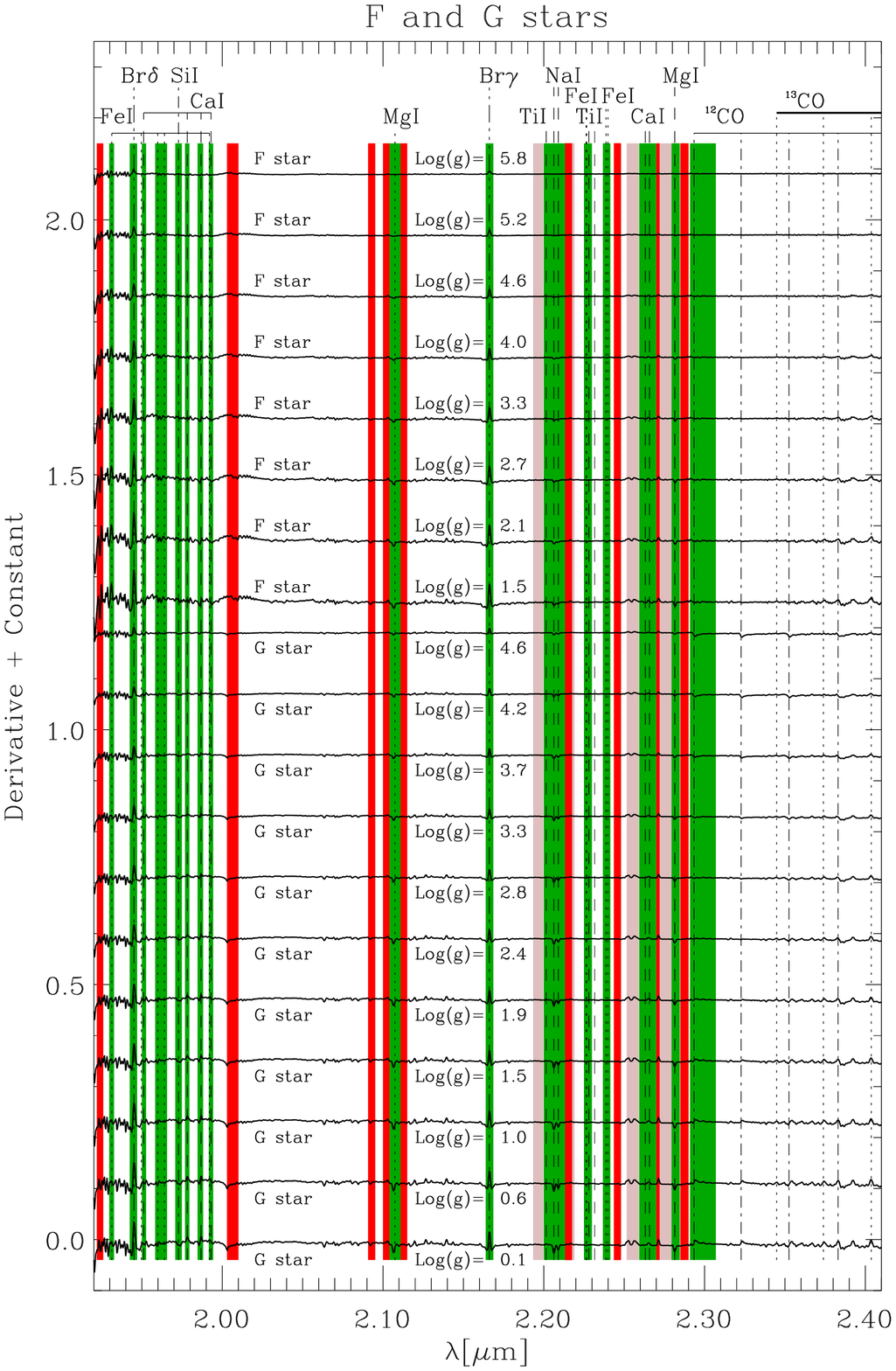}
\caption{$K$-band {\it sensitivity map} for surface gravity of F
    (top) and G-type stars (bottom). The sensitivity map for
    different gravity values is offset for displaying purposes and
    the central values of the corresponding $\log\,(g)$ bins are
    given. Symbols are as in Fig.\,\ref{fig:SupGian_SpT_K}.}
\label{fig:FG_Logg_K}
\end{figure*}

\begin{figure*}
\includegraphics[width=16truecm,angle=0]{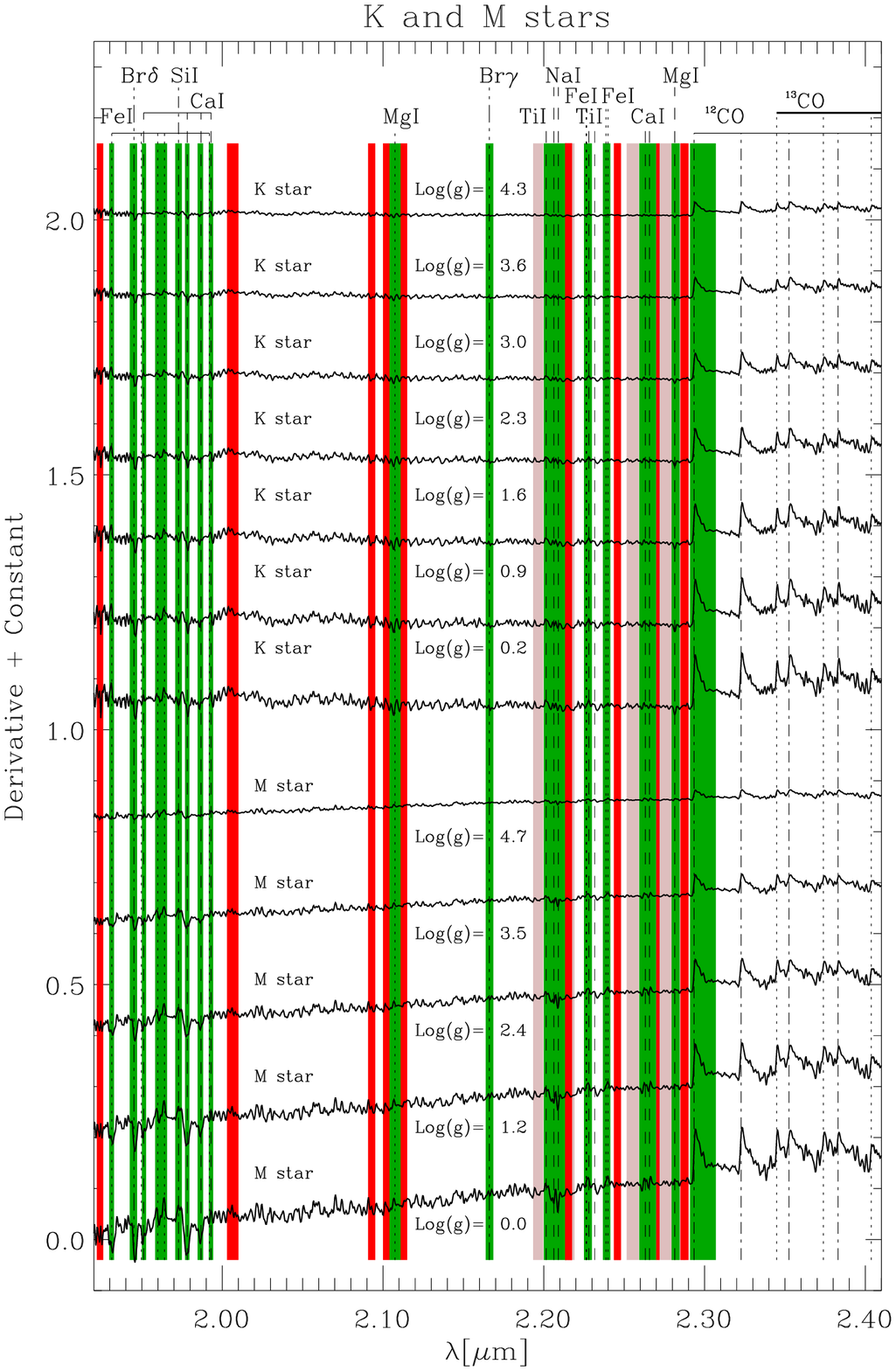}
\caption{As in Fig.\,\ref{fig:FG_Logg_K} but for K (top) and M-type stars
    (bottom).}
\label{fig:KM_Logg_K}
\end{figure*}

The $K$-band sensitivity map as a function of surface gravity is
plotted in Figs.\,\ref{fig:FG_Logg_K} and \ref{fig:KM_Logg_K} for the
F and G stars and the K and M stars, respectively. The sample is
divided according the SpT and the L stars are excluded, as done in the
$I$ band. The peaks at the Br$\delta$ and Br$\gamma$ lines are the
most prominent features for the F and G stars. They indicate that flux
at the core of the lines increases faster than the continuum flux with
increasing gravity, i.e., the lines become weaker for more compact
stars.  Mg, and to lesser extent Na, follow opposite trends, as
indicated by the small dips at these lines. The H{~\sc i} lines nearly
disappear in the K and M stars and the peaks corresponding to the CO
bandheads become the strongest features in the sensitivity map, being
characterized by a strong decrease towards lower surface gravity. The
Ca and Fe lines in the range $1.95-1.99\,\mu$m show small dips which
indicate a mild increase with increasing $\log\,(g)$. On the contrary,
the prominent Na and Ca doublets appear insensitive to the stellar
gravity.

\subsection{Definition of new $K$-band indices}
\label{sec:ind_def_K}

The sensitivity map allowed us to identify 16 $K$-band features that
could be used as indicators of the SpT and/or $\log\,(g)$. Eight
of them, falling in the spectral range $\lambda$=2.10$-$2.30\,$\mu$m,
correspond to absorption lines already studied. We followed the
previous definition of the line-strength indices by \citet{Iva04} for
Mg{~\sc i} at 2.11$\,\mu$m and Br$\gamma$, by \citet{Sil08} for
Fe{~\sc i} at 2.23$\,\mu$m, Fe{~\sc i} at 2.24$\,\mu$m, and Mg{~\sc i}
at 2.28$\,\mu$m, and by \citet{Ces09} for the Ca{~\sc i} and Na{~\sc
  i} doublets and the $^{12}$CO absorption band at 2.29$\,\mu$m. For
some of them in literature are present multiple index definitions. In
this case we preferred to use the index definition more suitable for
extragalactic studies. This is the case of Ca{~\sc i}, Na{~\sc i}, and
$^{12}$CO. The features for $\lambda>$2.29$\,\mu$m are not considered
in the analysis due to the difficulty of defining a reliable continuum
on the red side of the features.  Eight indices, falling in the
wavelength range $\lambda$=1.92$-$2.01\,$\mu$m are defined by adopting
the common continuum passband. The estimated continuum in the
wavelength range of the absorption features is shown with a straight
line in Figs.\,\ref{fig:SGiant_fitted_K}, \ref{fig:Giant_fitted_K},
and \ref{fig:Dwarf_fitted_K}. All the above definitions are listed in
Table\,\ref{tab:index_def_K}.  The indices were measured for all the
sample stars and their EWs are listed in
Tables\,\ref{tab:index_mis_K_1} and \ref{tab:index_mis_K_2}.

\begin{table}[t]
\caption{Definition of the bandpasses of the $K$-band indices.}
\label{tab:index_def_K} 
\begin{tabular}{@{}l@{ }c@{ }c@{ }c@{ }c@{}}
\hline
\hline
Index       & Element         & Central bandpass & Continuum  bandpasses &       Ref.\\
            &                 & ($\mu$m)         & ($\mu$m)              &           \\
\hline
Fe1         &\,Fe{~\sc i}     &\,1.9297--1.9327~&\,1.9220--1.9260, 2.0030--2.0100&(1)\\
Br$\delta$  &\,H{~\sc i} (n=4)&\,1.9425--1.9470~&\,1.9220--1.9260, 2.0030--2.0100&(1)\\
Ca1         &\,Ca{~\sc i}     &\,1.9500--1.9526~&\,1.9220--1.9260, 2.0030--2.0100&(1)\\
Fe23        &\,Fe{~\sc i}     &\,1.9583--1.9656~&\,1.9220--1.9260, 2.0030--2.0100&(1)\\
Si          &\,Si{~\sc i}     &\,1.9708--1.9748~&\,1.9220--1.9260, 2.0030--2.0100&(1)\\
Ca2         &\,Ca{~\sc i}     &\,1.9769--1.9795~&\,1.9220--1.9260, 2.0030--2.0100&(1)\\
Ca3         &\,Ca{~\sc i}     &\,1.9847--1.9881~&\,1.9220--1.9260, 2.0030--2.0100&(1)\\
Ca4         &\,Ca{~\sc i}     &\,1.9917--1.9943~&\,1.9220--1.9260, 2.0030--2.0100&(1)\\
Mg1         &\,Mg{~\sc i}     &\,2.1040--2.1110~&\,2.1000--2.1040, 2.1110--2.1150&(2)\\
Br$\gamma$  &\,H{~\sc i} (n=4)&\,2.1639--2.1686~&\,2.0907--2.0951, 2.2873--2.2900&(2)\\
Na$_{\rm d}$&\,Na{~\sc i}     &\,2.2000--2.2140~&\,2.1934--2.1996, 2.2150--2.2190&(3)\\
FeA         &\,Fe{~\sc i}     &\,2.2250--2.2299~&\,2.2133--2.2176, 2.2437--2.2479&(4)\\
FeB         &\,Fe{~\sc i}     &\,2.2368--2.2414~&\,2.2133--2.2176, 2.2437--2.2479&(4)\\
Ca$_{\rm d}$&\,Ca{~\sc i}     &\,2.2594--2.2700~&\,2.2516--2.2590, 2.2716--2.2888&(3)\\
Mg2         &\,Mg{~\sc i}     &\,2.2795--2.2845~&\,2.2700--2.2720, 2.2850--2.2874&(4)\\
$^{12}$CO   &\,$^{12}$CO(2,0) &\,2.2910--2.3070~&\,2.2516--2.2590, 2.2716--2.2888&(3)\\
\hline
\end{tabular}
References: (1) This paper, (2) \citet{Iva04}, (3) \citet{Ces09}, and (4) \citet{Sil08}.
\end{table}


\section{Spectral diagnostics in the $K$ band}
\label{sec:SpT_diag_K}

\subsection{Spectral diagnostics for the spectral type}

The new $K$-band indices are plotted as a function of spectral type in
Fig.\,\ref{fig:Ind_Spt_earlyK}. All indices within the
$1.92-2.01\,\mu$m range suffer from large scatter -- not surprising,
given their weakness and the poor atmospheric transmission in this
spectral region -- but a general increase of most Ca and Fe lines
towards later type stars is observed. The Ca2 and Ca3 indices have a
turnover at early- to mid-M stars, but their scatter is
significant. The Br$\delta$ index decreases in hotter stars stars
until hitting a turnover at early- to mid-K stars.

The Mg lines increase with SpT from F to K, but for later-type stars
the trend flattens or even reverts, and dwarf stars show abnormal
scatter. The Na, Ca, and Fe indices follow similar pattern but the
strength of the lines makes it more pronounced, especially the
turnover for K8-M0 supergiants. The Br$\gamma$ index decreases nearly
monotonically towards later SpTs. The $^{12}$CO index is insensitive
to SpT for F and early-G stars and it increases for later
SpTs. Moreover, it is also able to distinguish between luminosity
classes: the index is progressively weaker ranging from supergiant, to
giant, and to dwarf stars. The same results were obtained in the
analysis of the spectral diagnostics for the $T_{\rm eff}$.

With the exception of a few features, the behavior of indices is
similar for stars of the three luminosity classes.

\subsection{Spectral diagnostics for surface gravity and metallicity}

\begin{figure*}
\includegraphics[width=9truecm,angle=0]{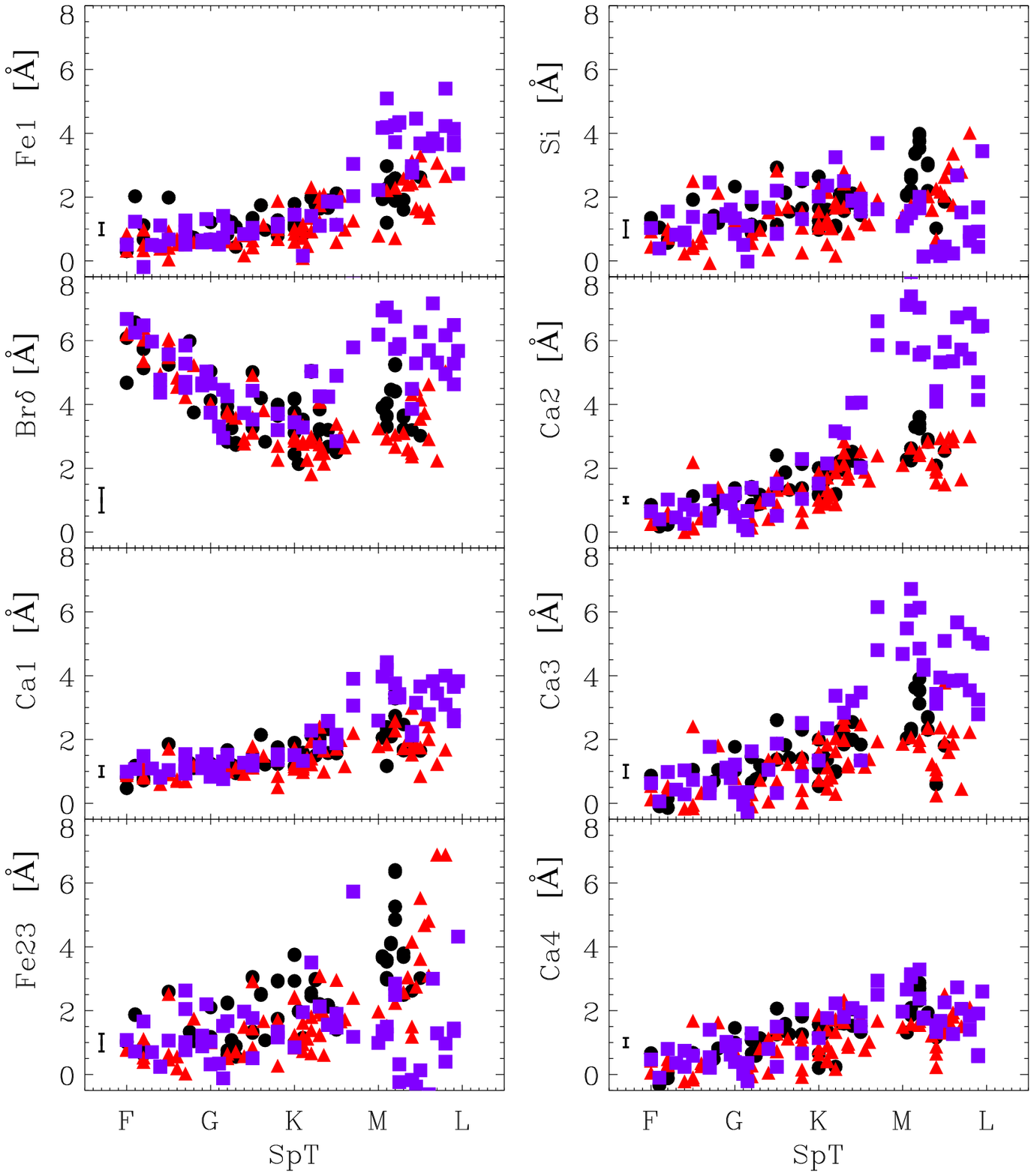}
\includegraphics[width=9truecm,angle=0]{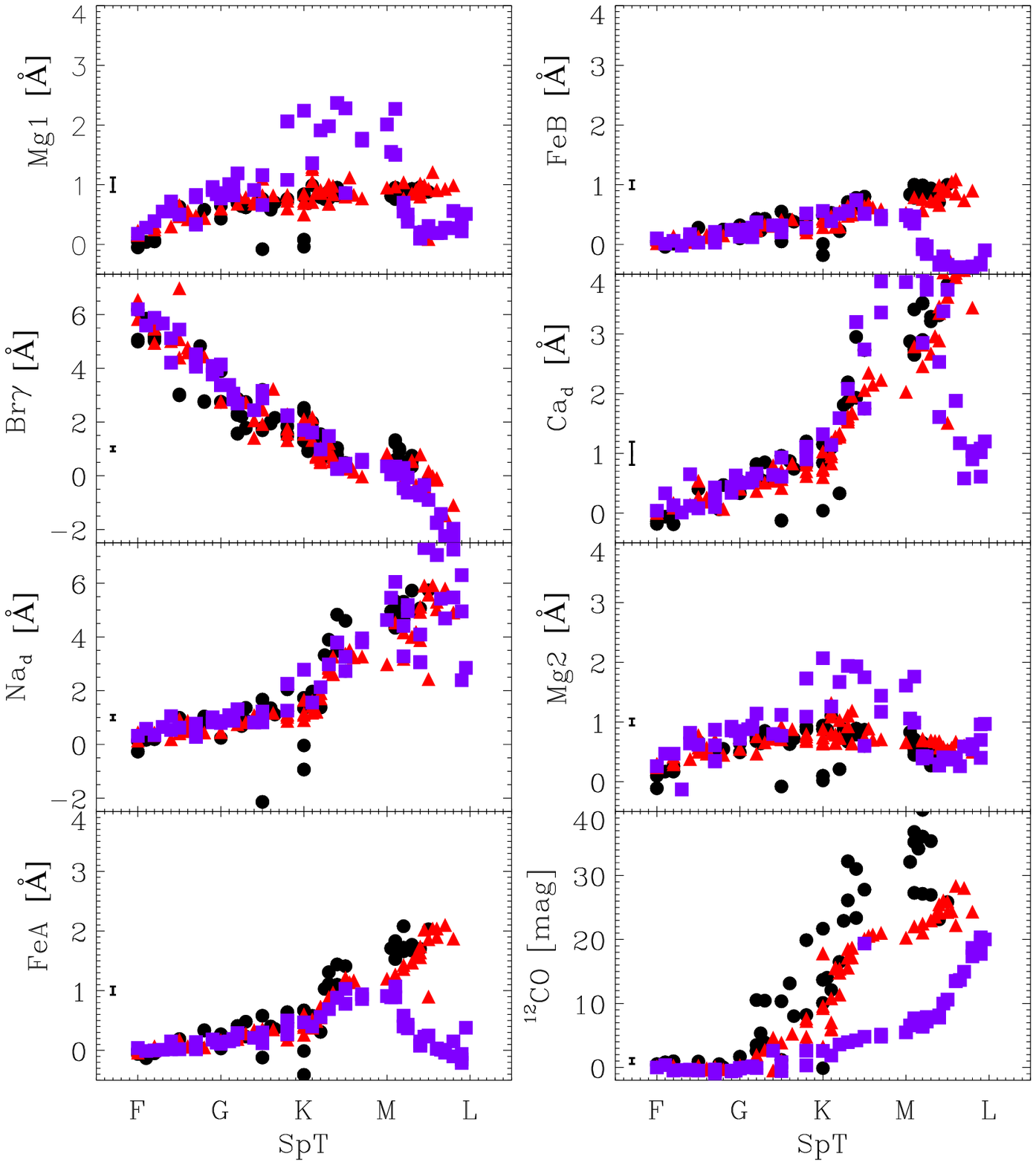}
\includegraphics[width=9truecm,angle=0]{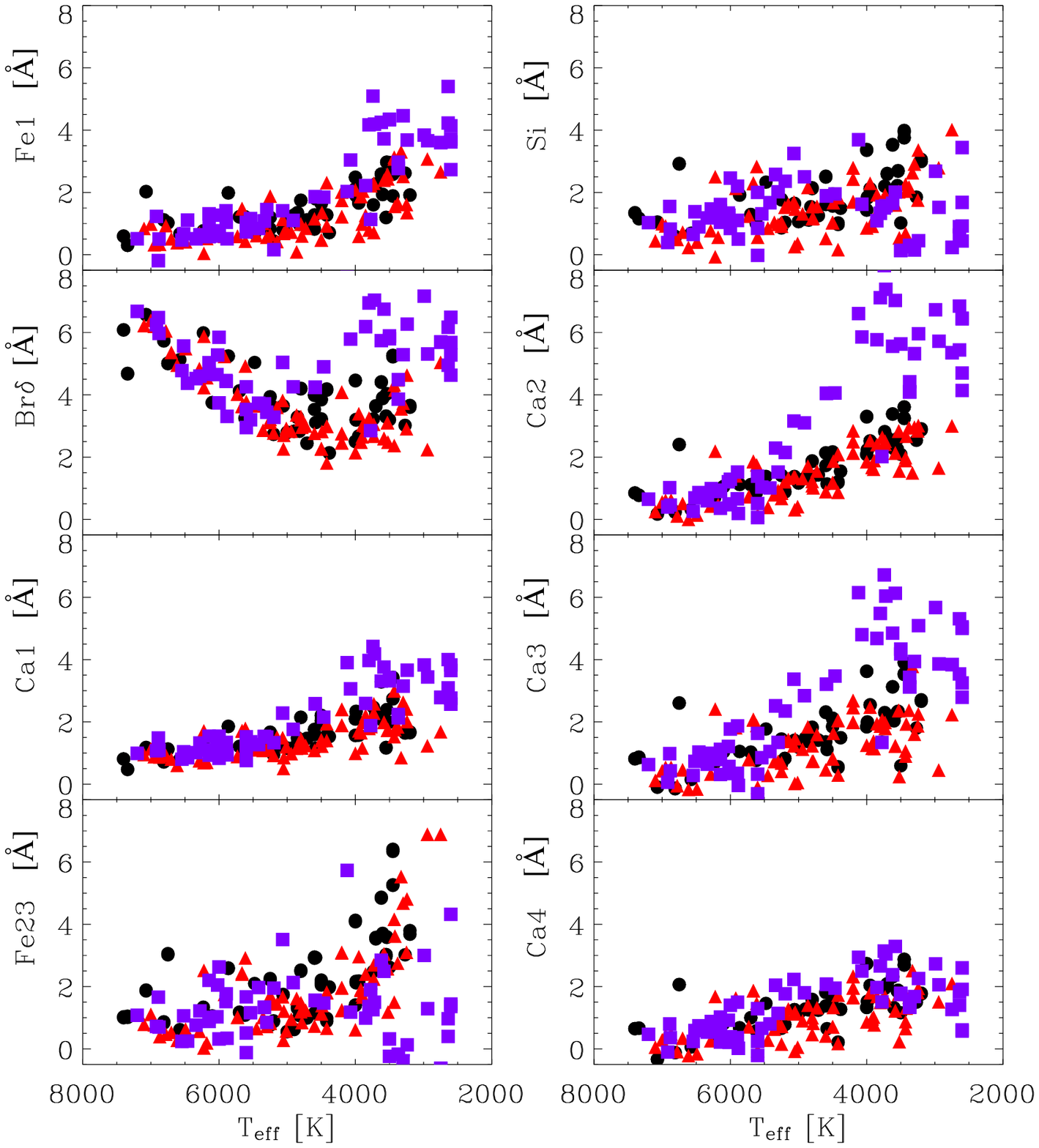}
\includegraphics[width=9truecm,angle=0]{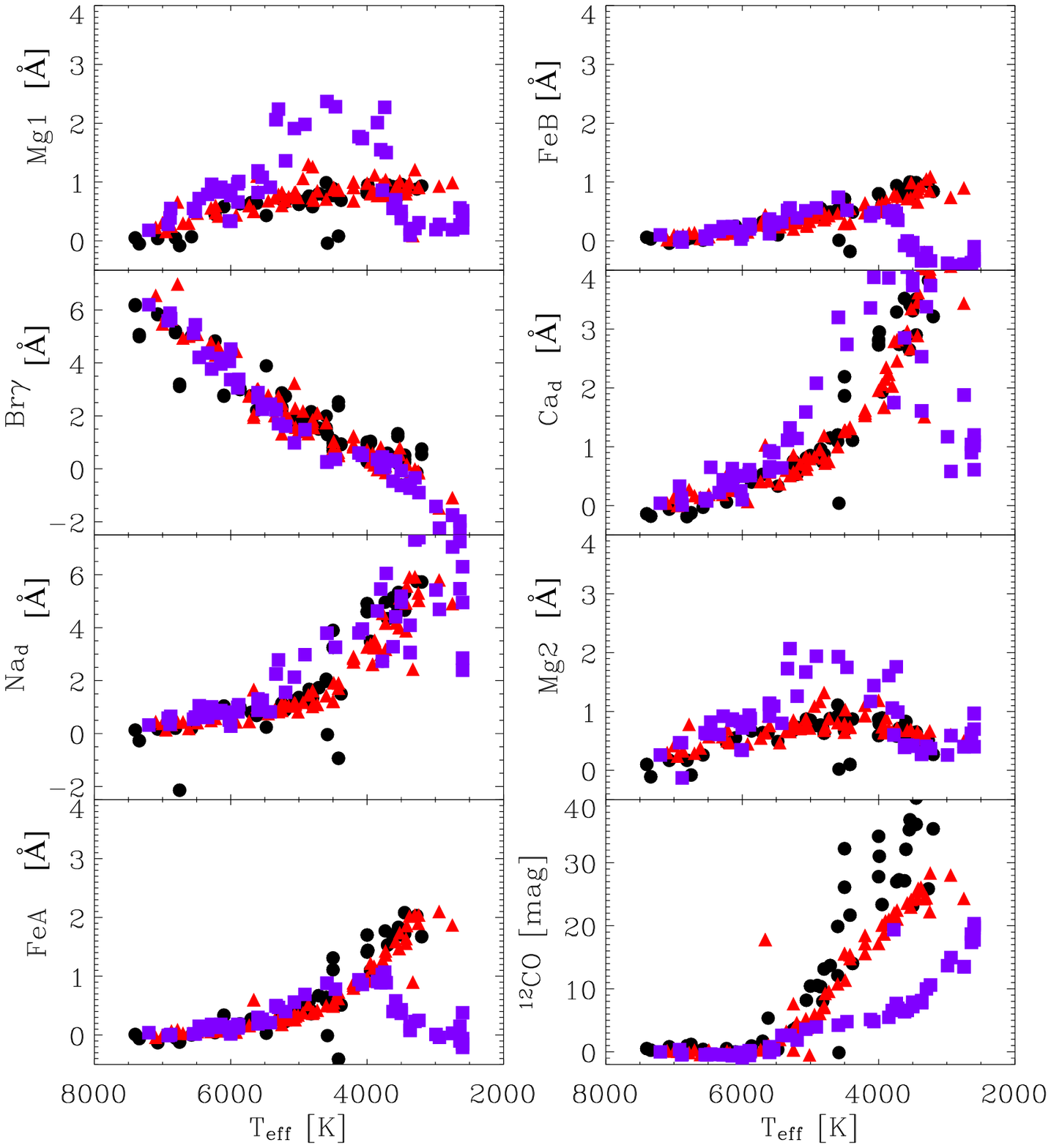}
\caption{As in Fig.\,\ref{fig:Ind_Spt} but for $K$-band indices.}
\label{fig:Ind_Spt_earlyK}
\end{figure*}

\begin{figure*}
\includegraphics[width=9truecm,angle=0]{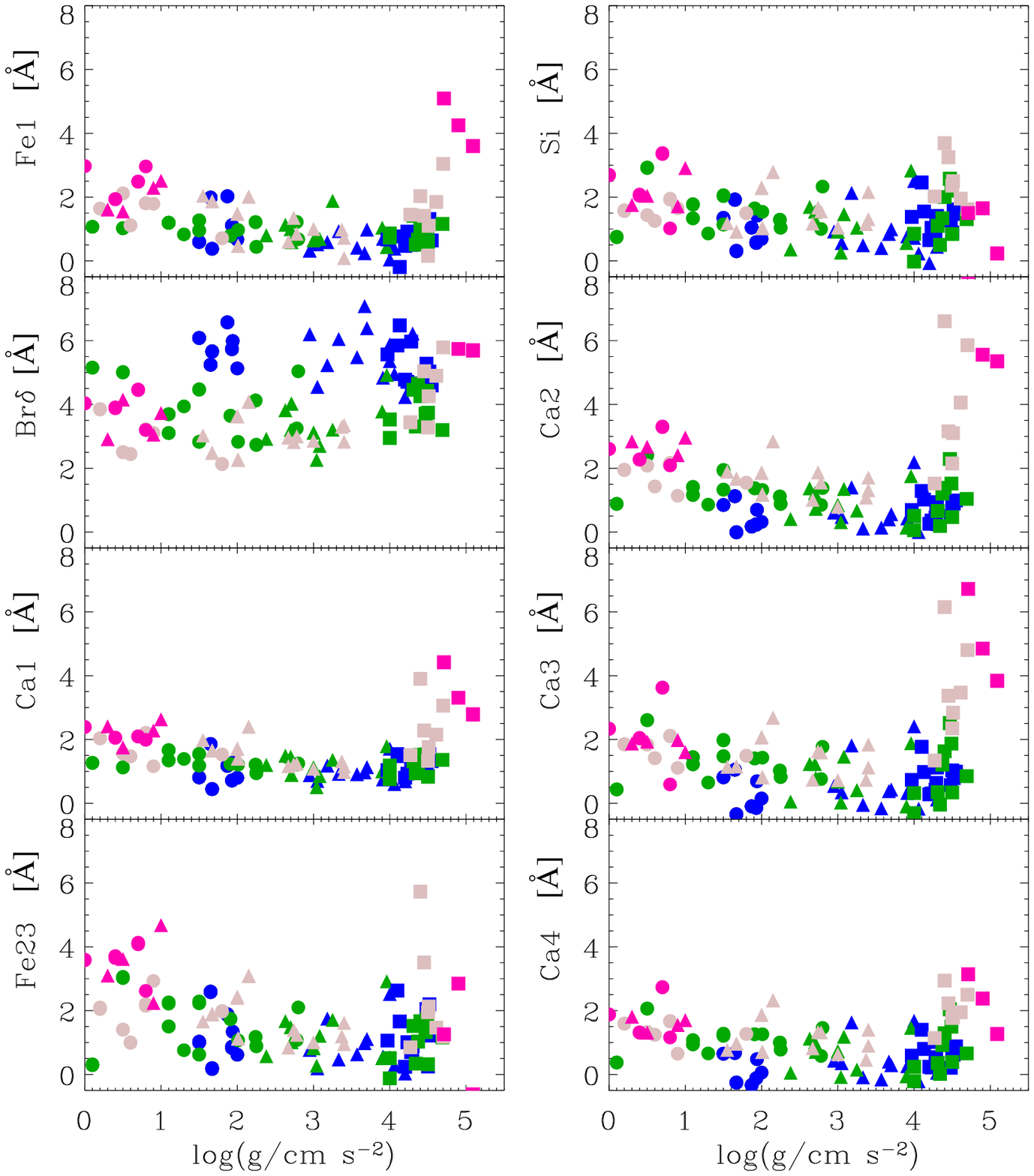}
\includegraphics[width=9truecm,angle=0]{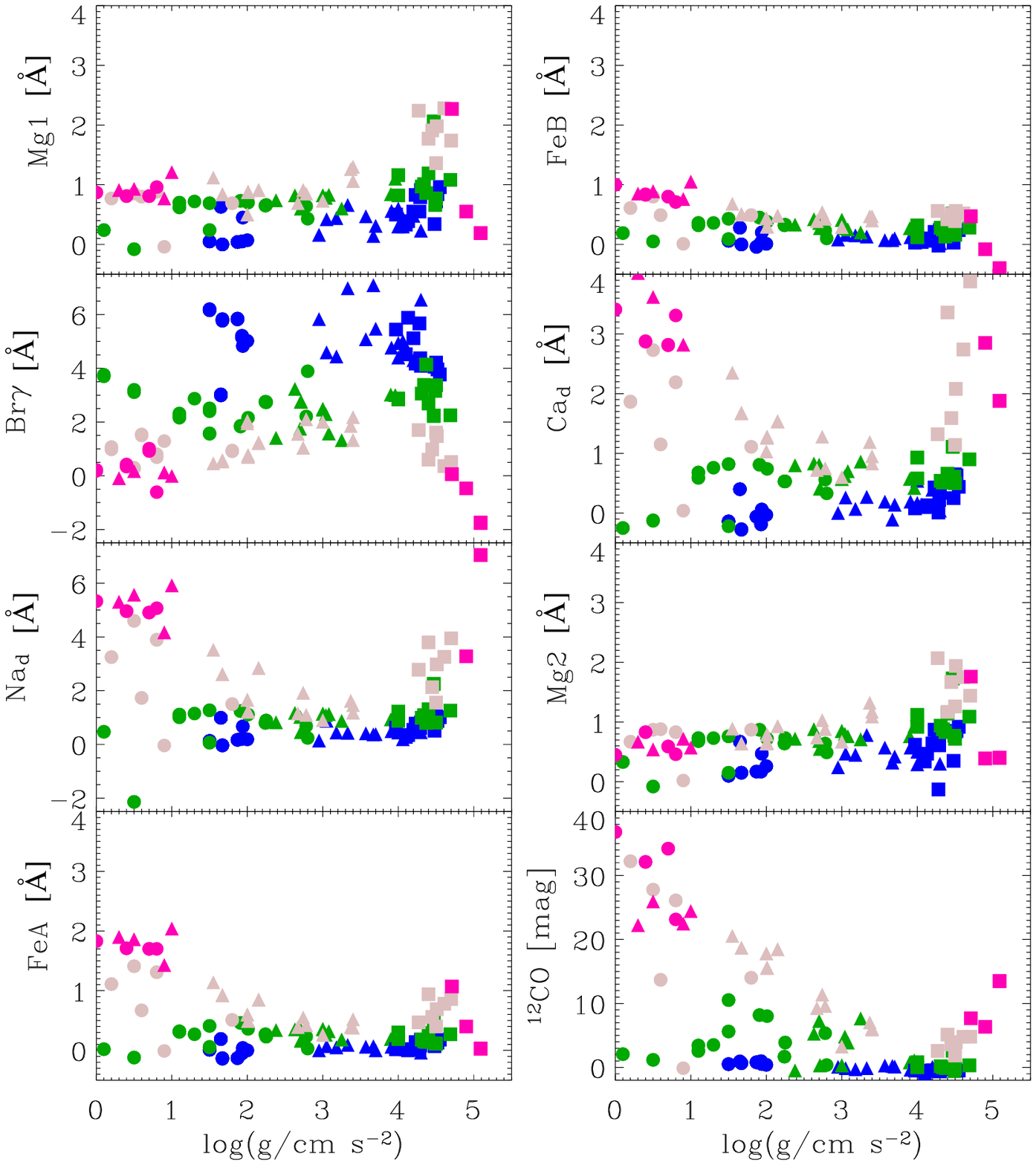}
\includegraphics[width=9truecm,angle=0]{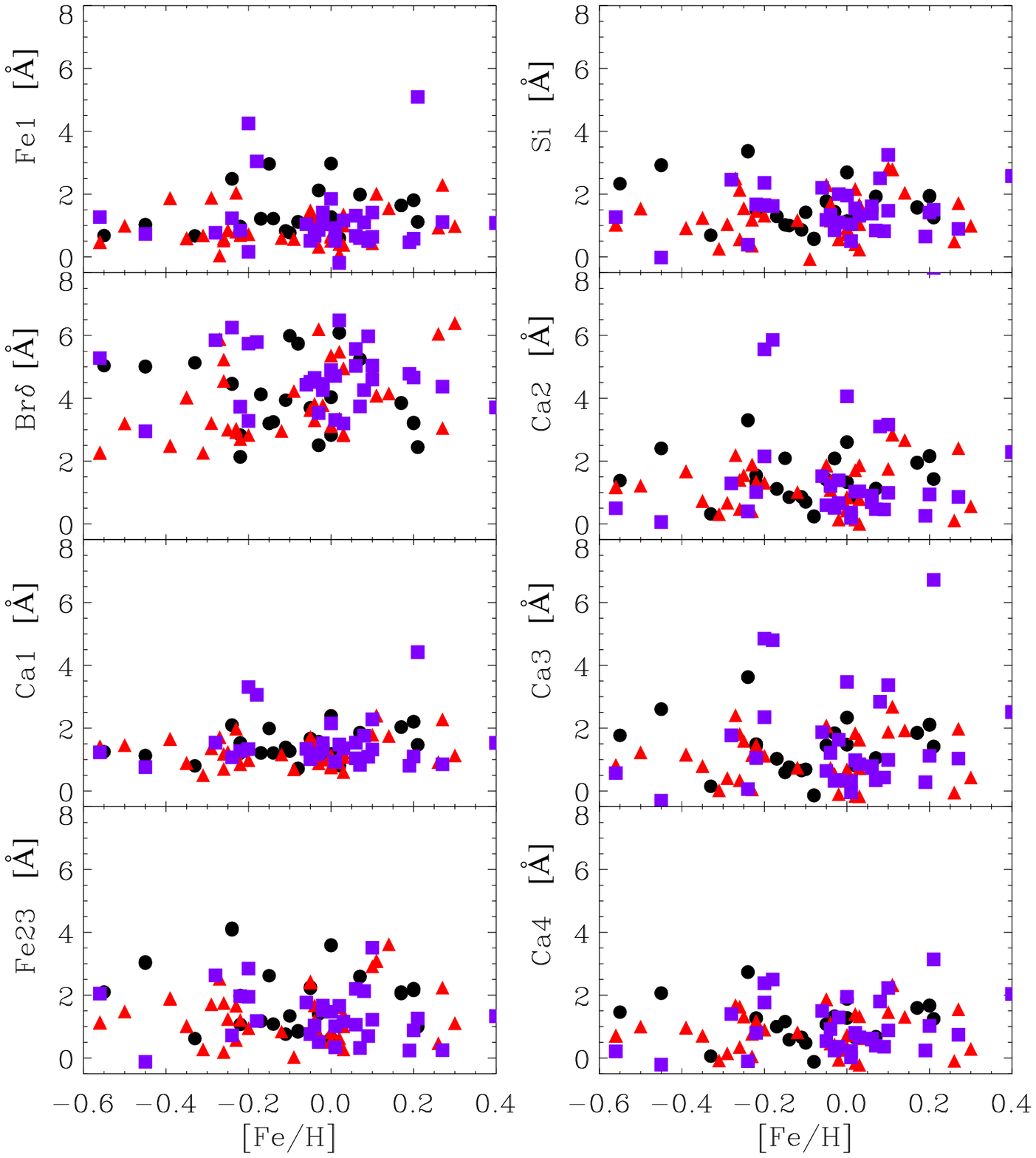}
\includegraphics[width=9truecm,angle=0]{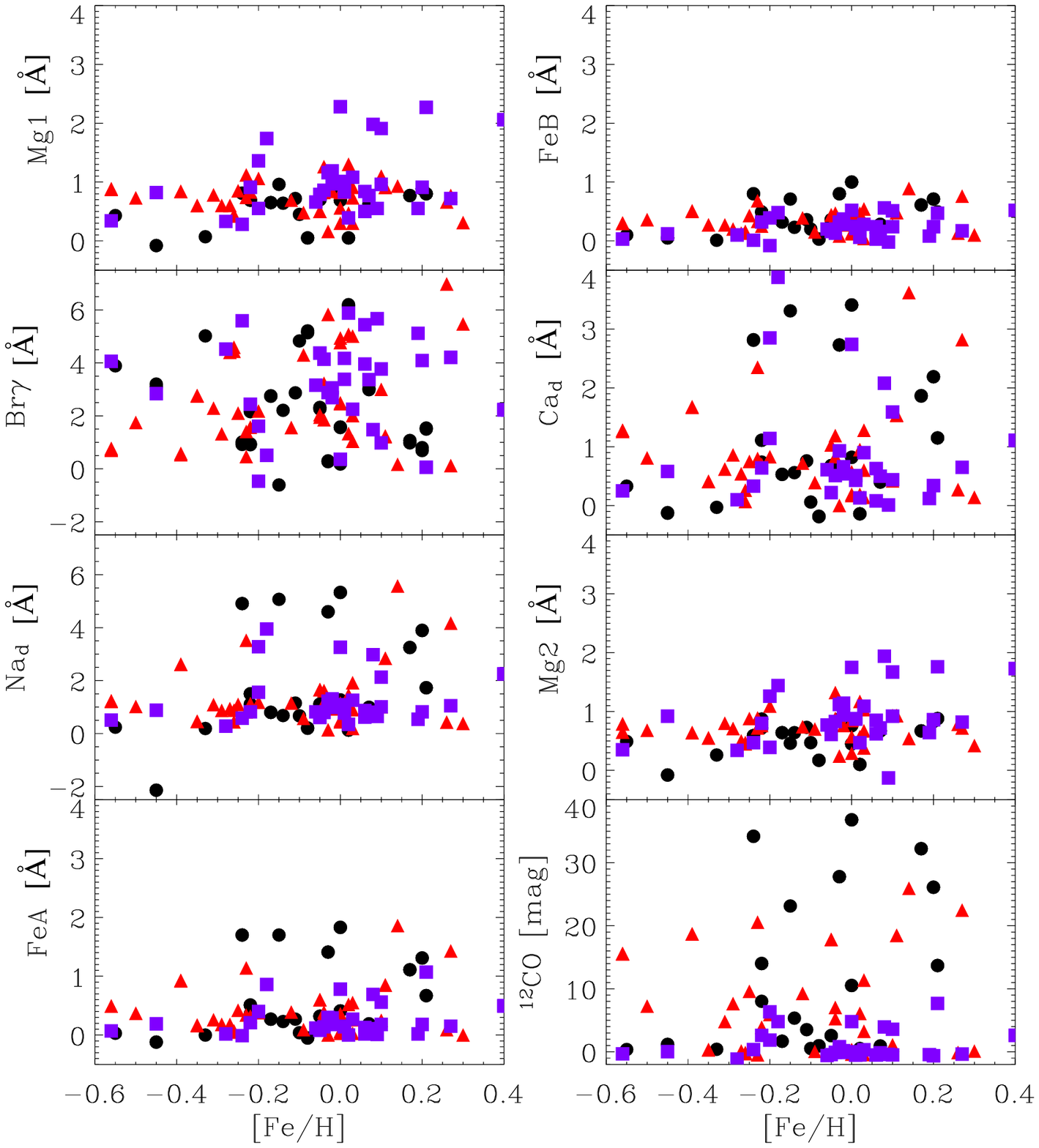}
\caption{As in Fig.\,\ref{fig:Ind_Gr} but for $K$-band indices.}
\label{fig:Ind_Gr_earlyK}
\end{figure*}

We also studied how the indices vary as function of the surface
gravity and metallicity (Fig.\,\ref{fig:Ind_Gr_earlyK}). Most features
show no significant changes within the errors with two exceptions. The
molecular CO band is sensitive to $\log (g)$, as expected. The metal
lines of K and M stars peak at the extremely high gravity. This effect
did not appear for the $I$-band indices and it is probably due to a
combination of two factors: in late-type dwarfs the $T_{\rm eff}$ and
$\log (g)$ are correlated (i.e., the start are ordered along the main
sequence) and the features do show temperature dependence, as
demonstrated in Fig.\,\ref{fig:Ind_Spt_earlyK}. The large scatter of
Na and Ca indices for low gravity K stars is puzzling.

Finally, no trend is observed with [Fe/H] for any of the investigated
lines, as expected for the narrow metallicity range of the sample.

\section{Discussion and conclusions}

\label{Sec:limits}

The \citet{Cen01b,Cen02} library, which is limited to the
$I$-band spectral region, is characterized by a larger number of
stars (706 stars with $T_{\rm eff}\,=\,3300-25000$\,K) compared to
and our sample (198 stars with $T_{\rm eff}\,=\,2600-7200$\,K) and
by a higher spectral resolution. This gives us the opportunity to
check if our results are affected by poor stellar parameter
sampling, spectral resolution, and differences in data reduction.

\begin{figure}
\includegraphics[width=9truecm,angle=0]{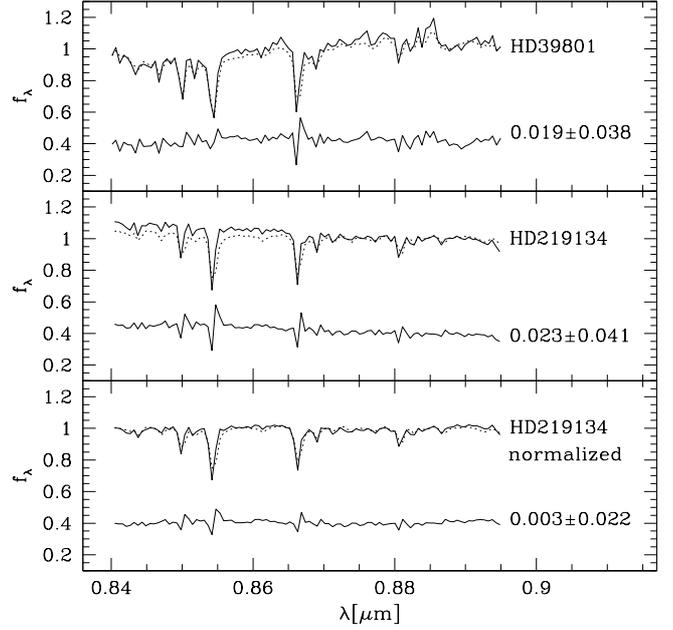}
\caption{Direct comparison of the spectra of Betelgeuse
    (HD\,39801, top panel) and HD\,219134 (middle panel) from the
    libraries of \citet[][solid lines]{Ray09} and \citet[][dotted
      lines]{Cen01b,Cen02}. In the bottom panel the continuum
    normalized spectra of HD\,219134 are shown. In each panel the
    difference of the two corresponding spectra are plotted (shifted
    by +0.4 for displaying purposes) and the rms value is given.}
\label{fig:LibCom}
\end{figure}

{\em Spectra direct comparison.\/} -- We compared directly spectra
from the IRTF and \citet{Cen01b,Cen02} libraries, rebinning the latter
to the resolution of the IRTF spectra. Two extreme cases were
considered: the M supergiant Betelgeuse (HD\,39801), one of the
brightest stars in both samples, and a flaring K3 dwarf HD\,219134. In
Fig.\,\ref{fig:LibCom} we show for each star the two spectra
superimposed and their difference.  In the case of HD\,219134 we
notice a slight difference in the continuum slope. We therefore
compared the two spectra also after the continuum normalization
(Fig.\,\ref{fig:LibCom} bottom panel). We see that no telluric
corrections residuals are present and the strength of the individual
spectral features are the same. We therefore are confident that the
IRTF and \citet{Cen01b,Cen02} libraries are consistent.

\begin{figure}
\includegraphics[width=9truecm,angle=0]{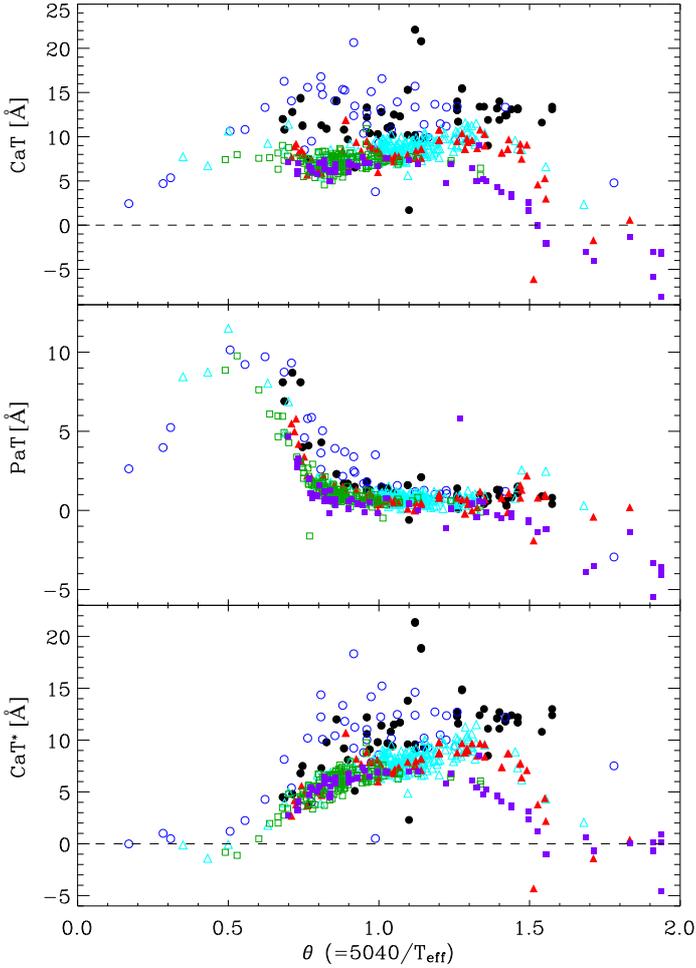}
\caption{EW of the CaT (top panel), PaT (middle panel), and CaT* index
(bottom panel) as a function of $\theta\,=\,5040/T_{\rm eff}$ for the
stars of the IRTF sample (filled symbols) and the sample by
\citet[][open symbols]{Cen02} sample.  The different symbols
correspond to the supergiant (stars), giant (triangles), and dwarf
stars (squares), respectively.}
\label{fig:CaT}
\end{figure}

{\em Calcium Triplet comparison.\/} -- Indices as the CaT were
extensively studied both with empiric that theoretic methods. A
comprehensive study was done by Cen01 that extensively studied the
behavior of this spectral feature in respect to $T_{\rm eff}$,
$\log\,(g)$, and [Fe/H].  We compare our sample CaT behavior with
Cen01 adopting their definition of the CaT* index that removes the
Paschen line contamination. The CaT* is defined as: $\rm
CaT^{\ast}\,=\,CaT-0.93\,\times\,PaT$, where the $\rm
CaT\,=\,Ca1+Ca2+Ca3$ and $\rm PaT\,=\,Pa1+Pa2+Pa3$ indices measure the
strength of the raw CaT and three Paschen lines. We limited the
comparison to stars within the same metallicity range as in
\citet{Cen02}.  Different luminosity classes were analyzed
separately. The comparison of the samples shows excellent
agreement and similar scatter (Fig.\,\ref{fig:CaT}).

{\em Sensitivity map comparison.\/} -- We tested if the library
limited number of stars and systematic errors in the stellar
parameters can affect the sensitivity maps. This has been done by
deriving the sensitivity maps with the \citet{Cen01b,Cen02}
library. The maps are remarkably similar.\\

We demonstrated that the study of derivatives yields qualitatively
consistent results with respect to the direct investigation of the
behaviour of line-strength indices. However, it is not possible to
predict the strength of a gradient and the intrinsic scatter of the
correlations.  The latter is due to the smoothing applied to the data
using a second-degree polynomial as fitting function in the derivation
of the model spectrum. In fact, the introduction of the fitting
function was necessary to fill gaps in the parametric space due to
irregular sampling and to smooth the data before perform its
derivative.

\begin{figure}
\includegraphics[width=9truecm,angle=0]{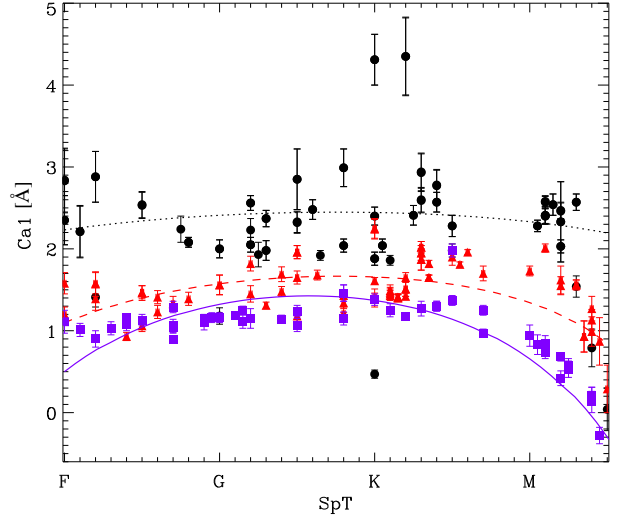}
\caption{Equivalent width of the Ca1 index as a function of spectral
    type for the observed and model spectrum of the sample stars. The
    different symbols correspond to the measurements in the observed
    spectra of the supergiant (circles), giant (triangles), and dwarf
    stars (squares), respectively. The different lines are for
    measurements in the model spectrum of of the supergiant (dotted
    line), giant (dashed line), and dwarf stars (solid line),
    respectively.}
\label{fig:Ind_Ca1Spt}
\end{figure}

To address this problem we measured the EWs on both the model and
observed spectra. Let's consider, as an example, the case of Ca1
dependence on SpT. The EWs of the observed spectra are characterized
by a significant scatter which is different for the three luminosity
classes, whereas the EWs measured on the model spectrum change smoothly
with SpT (Fig.\,\ref{fig:Ind_Ca1Spt}).  This means that, although for
each wavelength an independent fit is applied along the SpT direction,
the different fits maintain a degree of coherence along the wavelength
direction such that a smooth trend of the EW is observed when it is
measured on the model spectrum. On the other hand the EWs of the model
spectrum do follow the trend measured on real data with different
degrees of goodness. In this respect, the choice of the degree of the
fitting function plays a relevant role. The introduction of a
higher-order polynomial generally allows a better description of the
data but the noise on the derivative increases severely limiting its
effectiveness in defining indices.  Our conclusion is that a
second-order polynomial leads to a sensitivity map that qualitatively
predict the variation of a spectral feature with respect to a given
physical parameter.  Therefore the sensitivity map is useful as an
objective tool to define the best bandpasses for the spectral feature
and their adjacent continuum, whereas the measurements on the real
spectra yield more reliable correlations between the EWs and the
stellar physical parameters. Last but not least, the index definitions
derived from the sensitivity map depend on the spectral resolution of
the library.\\

{\em Sensitivity of indices to velocity dispersion broadening.\/}
-- A straightforward application of the indices system is the study
of the unresolved stellar population in galaxies. It is therefore
important to investigate how our measurements are affected by the
unavoidable velocity broadening due to the internal galaxy
kinematics.

\begin{figure*}
\includegraphics[height=18truecm,angle=90]{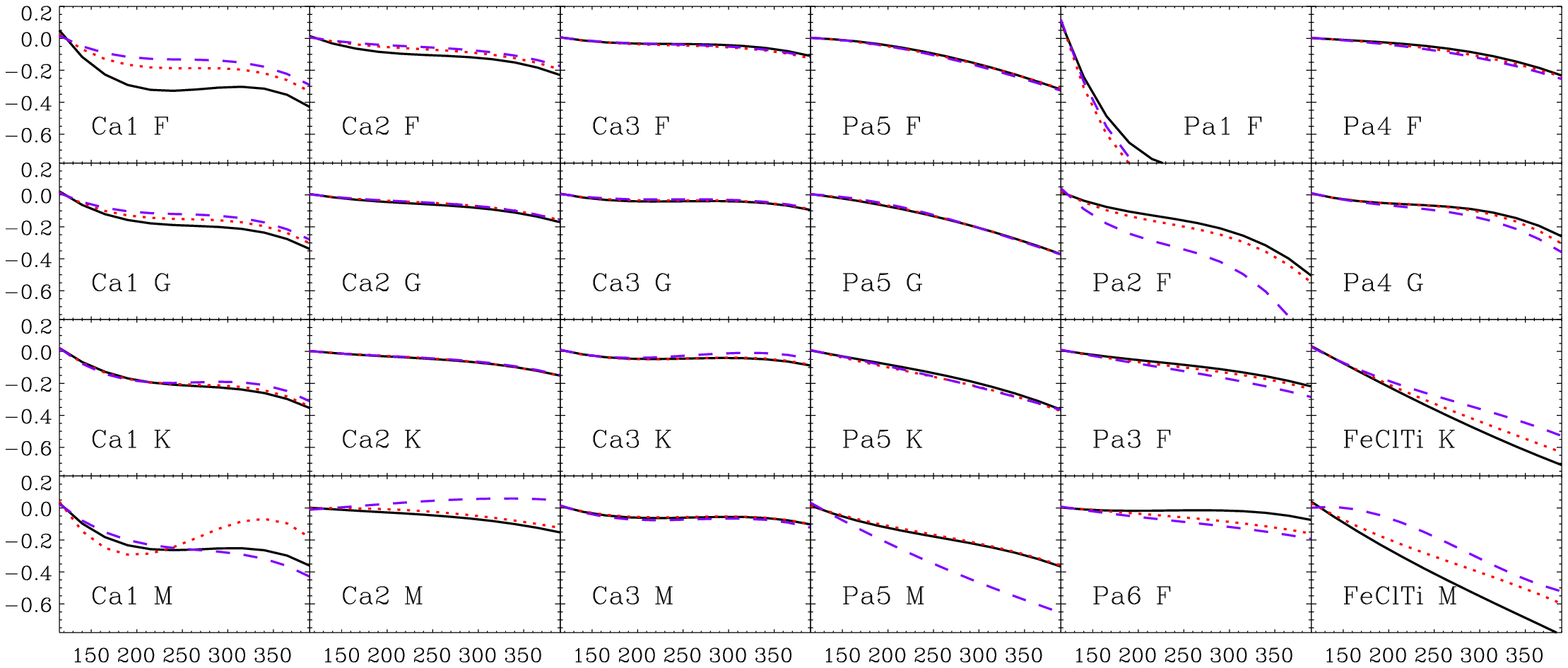}
\caption{$I$-band indices sensitivity to the velocity dispersion
    broadening. The indices relative variations for different
    luminosity classes and spectral types are shown: supergiants
    (solid line), giants (dashed line) and dwarfs (dotted line).}
\label{fig:VelDisp_I}
\end{figure*}

\begin{figure*}
\includegraphics[height=18truecm,angle=90]{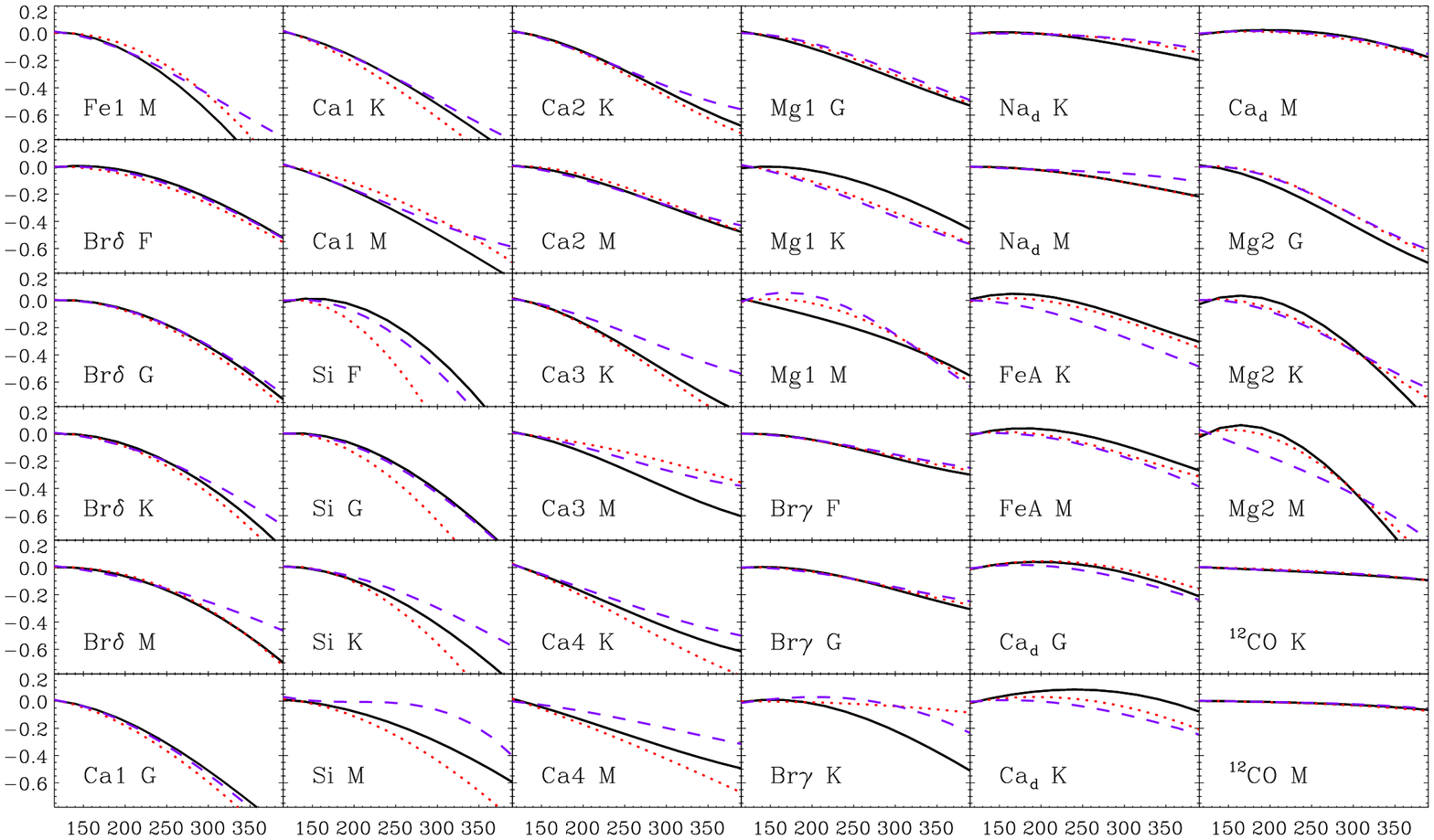}
\caption{ As in Fig. \ref{fig:VelDisp_I} but for the $K$-band indices.}
\label{fig:VelDisp_K}
\end{figure*}

We broadened all the spectra by convolving them with a Gaussian of
$\sigma$ varying from 115 to 400 km s$^{-1}$, in steps of 25 km
s$^{-1}$. The indices were measured for all the broadened spectra
and a third-order polynomial fit was done to the relative changes of
each index value as a function of velocity dispersion. Figures
\ref{fig:VelDisp_I} and \ref{fig:VelDisp_K} illustrate the
$\Delta$(Index)/Index($\sigma_0$) values for the different
luminosity classes and spectral types. It has to be noted that some
indices have a low EW value, typically $\lesssim 1$ \AA \quad for
some of the spectral types considered. The EW of these indices shows
a large variation even for small values of $\sigma$ and goes to zero
for $\sigma > 200-300$ km s$^{-1}$. The faint indices are not shown
in Fig. \ref{fig:VelDisp_I} and \ref{fig:VelDisp_K}. The analysis
shows that there is a very strong correlation between the line
strength and the effect of the broadening, being the strongest lines
less affected. This is true in both bands. The $K$-band indices are
generally more affected by broadening due to the intrinsic faintness
of the lines.\\


This work extended the previously defined sensitivity indices
\citep{Wor94} to a broader concept of sensitivity map and tested if
they allows us to identify the spectral features that can be used as a
proxy for $T_{\rm eff}$, $\log\,(g)$, and [Fe/H]. This empirical
method is applied to $I$- and $K$-band absorption features for the
star spectra of the IRTF library \citep{Cus05,Ray09}. The main
results are:

\begin{itemize}

\item sensitivity map allows to fine tune the best definition
  for the line-strength indices (i.e., the bandpass limits for the
  line and nearby continuum);

\item sensitivity map reliably predicts the variation of a
spectral feature with respect to a given physical parameter but not
its absolute strength;

\item spectral line blends are obvious on sensitivity map when
  the blended features are characterized by a different behavior with
  respect to some physical stellar parameters;

\item the EWs of new indices were measured for the IRTF star sample,
  and they will be useful for stellar population synthesis models and
  can be used to get element-by-element abundances for unresolved
  stellar population studies in galaxies;

\item a systematic search for reliable $T_{\rm eff}$, $\log\,(g)$, and
  [Fe/H] for the IRTF sample stars was carried out and the available
  physical parameters are reported.

\end{itemize}

We develop a fast and efficient method to identify those features that
are sensible to different physical stellar parameters. The method is
free from any assumption and it can be applied to any star or
line-strength index. In a forthcoming paper we will extend such an
analysis to $Y$, $J$, $H$, and $L$ bands to define new indices. A
straightforward next step is the extensive use of the sensitivity map
in the upcoming X-Shooter Spectral Library \citep[XLS,][]{Che11}.


\begin{acknowledgements}       
M.C. acknowledges the European Southern Observatory for hospitality at
Santiago Headquarters while this paper was in progress. This work was
supported by Padua University through the grants CPDA089220/08,
60A02-5934/09, and 60A02-1283/10 and by Italian Space Agency through
the grant ASI-INAF I/009/10/0. M.C acknowledges financial support from
Padua University grant CPDR095001/09 and
CPDR115539/11. L.M. acknowledges financial support from Padua
University grant CPS0204.

\end{acknowledgements}         

\bibliographystyle{aa}
\bibliography{IRTF}

\clearpage

Tables \ref{tab:sample_params}, \ref{tab:index_mis_I_1}, \ref{tab:index_mis_I_2},
\ref{tab:index_mis_K_1} and \ref{tab:index_mis_K_2} that follow are an
important part of the project, but they are placed here 
because they are large.

\onecolumn
\begin{center}\small
\begin{longtable}{ l l l l l l r r r }
\caption[IRTF library]{Physical parameters of the stars of the IRTF
library.}
\label{tab:sample_params}\\
\hline
\hline
Star IDs  & R.A. \, \, \, \, \, \, \, \, Dec. &Sp.\,Type~~~~~~&$M_V$   &~$T_{\rm eff}$ &~$\log\,(g)$~& [Fe/H]~~& Parallax & Ref.\\
          & (J2000)\, \, \, \, \, \,(J2000)   &               & mag    &   ~K          &             &         & mas      &     \\
 (1)      & (2)\, \, \, \, \, \, \, \,\, \, \,(3) & (4)       & (5)    & (6)           & (7)         & (8)     & (9)      & (10)\\
\hline                                                  
\endfirsthead
\multicolumn{3}{c}%
{{\bfseries \tablename\ \thetable{} -- continued from previous page}} \\
\hline
\hline
Star IDs  & R.A. \, \, \, \, \, \, \, \, Dec. &Sp.\,Type~~~~~~&$M_V$   &~$T_{\rm eff}$ &~$\log\,(g)$~& [Fe/H]~~& Parallax & Ref.\\
          & (J2000)\, \, \, \, \, \,(J2000)   &               & mag    &   ~K          &             &         & mas      &     \\
 (1)      & (2)\, \, \, \, \, \, \, \,\, \, \,(3) & (4)       & (5)    & (6)           & (7)         & (8)     & (9)      & (10)\\
\hline                                                  
\endhead
HD007927         &  01 20 04.9 \, \,   $+$58 13 53.8    & F0Ia       &    5.01 &    7341 &      -- &       -- &     1.40  &   1 \\
HD135153         &  15 14 37.3 \, \,   $-$31 31 08.8    & F1II       &    4.92 &    7070 &    1.87 &       -- &     2.86  &   2 \\
HD006130         &  01 03 37.0 \, \,   $+$61 04 29.3    & F0II       &    5.91 &    7400 &    1.50 &     0.02 &     2.16  &   3 \\
HD089025         &  10 16 41.4 \, \,   $+$23 25 02.3    & F0III      &    3.44 &    6950 &    2.95 &  $-$0.03 &    12.56  &   2 \\
HD013174         &  02 09 25.3 \, \,   $+$25 56 23.6    & F2III      &    4.98 &    7000 &    3.7  &     0.3  &    10.19  &   4 \\
HD027397         &  04 19 57.7 \, \,   $+$14 02 06.7    & F0IV       &    5.56 &    7100 &    4.3  &       -- &    22.31  &   5 \\
HD108519         &  12 27 46.3 \, \,   $+$27 25 21.9    & F0Vn       &      -- &    7200*&      -- &       -- &     6.90  &  -- \\
HD173638         &  18 46 43.3 \, \,   $-$10 07 30.1    & F2Ib-II    &    5.73 &    7090 &    1.67 &  $-$0.10 &  $-$0.07  &   2 \\
\end{longtable}
\end{center}

\begin{center}\small
\begin{longtable}{lrrrrrrr}
\caption{EW of the $I$-band indices. Part 1.}
\label{tab:index_mis_I_1} \\

\hline
\hline
Star IDs  & Ca1   & Ca2   & Ca3   &   Pa1   &   Pa2   &   Pa3   &  Mg  \\
\hline                                                 
\endfirsthead
\multicolumn{8}{c}%
{{\bfseries \tablename\ \thetable{} -- continued from previous page}} \\
\hline
\hline
Star IDs  &  Ca1   & Ca2   & Ca3   &   Pa1   &   Pa2   &   Pa3   &  Mg  \\
\hline                                                  
\endhead
\hline
\\
\multicolumn{8}{c}{\textbf{Supergiants}}\\
HD007927  &~  2.83  $\pm$   0.20  &~   3.72  $\pm$   0.16  &~   4.12  $\pm$   0.09  &~   1.48  $\pm$   0.12  &~   2.76  $\pm$   0.21  &~   2.85  $\pm$   0.15  &~   0.11  $\pm$   0.06    \\
HD006130  &~  2.35  $\pm$   0.15  &~   4.50  $\pm$   0.12  &~   5.00  $\pm$   0.06  &~   1.08  $\pm$   0.09  &~   2.97  $\pm$   0.16  &~   4.14  $\pm$   0.11  &~   0.17  $\pm$   0.03    \\
HD135153  &~  2.21  $\pm$   0.16  &~   4.69  $\pm$   0.12  &~   5.67  $\pm$   0.05  &~   1.10  $\pm$   0.10  &~   3.39  $\pm$   0.16  &~   4.12  $\pm$   0.09  &~   0.08  $\pm$   0.07    \\
HD164136  &~  1.41  $\pm$   0.06  &~   2.99  $\pm$   0.05  &~   3.13  $\pm$   0.04  &~   0.45  $\pm$   0.04  &~   1.35  $\pm$   0.07  &~   2.38  $\pm$   0.08  &~   0.17  $\pm$   0.04    \\
HD182835  &~  2.88  $\pm$   0.16  &~   5.44  $\pm$   0.13  &~   5.92  $\pm$   0.07  &~   1.30  $\pm$   0.10  &~   3.16  $\pm$   0.20  &~   3.74  $\pm$   0.11  &~   0.18  $\pm$   0.05    \\
HD213306  &~  2.54  $\pm$   0.08  &~   6.53  $\pm$   0.06  &~   5.16  $\pm$   0.07  &~   0.50  $\pm$   0.05  &~   0.82  $\pm$   0.17  &~   1.08  $\pm$   0.12  &~   0.28  $\pm$   0.06    \\
\hline
\end{longtable}
\end{center}

\begin{center}\small
\begin{longtable}{lrrrrrrr}
\caption{EW of the $I$-band indices. Part 2.}
\label{tab:index_mis_I_2} \\

\hline
\hline
Star IDs  &  Pa3  &  Pa4  &  Pa5  &  Pa6  &  CaT  &  PaT  &  CaT* \\
\hline                                                 
\endfirsthead
\multicolumn{8}{c}%
{{\bfseries \tablename\ \thetable{} -- continued from previous page}} \\
\hline
\hline
Star IDs  &  Pa3  &  Pa4  &  Pa5  &  Pa6  &  CaT  &  PaT  &  CaT* \\
\hline                                                  
\endhead
\hline
\\
\multicolumn{8}{c}{\textbf{Supergiants}}\\
HD007927  &~   3.48  $\pm$   0.06  &~   3.16  $\pm$   0.04  &~   2.02  $\pm$   0.02  &~   3.09  $\pm$   0.35  &~   10.8  $\pm$   0.9   &~   6.9  $\pm$   0.4  &~    4.4  $\pm$   0.8 \\
HD006130  &~   4.18  $\pm$   0.12  &~   4.13  $\pm$   0.04  &~   1.16  $\pm$   0.02  &~   4.86  $\pm$   0.17  &~   12.0  $\pm$   0.6   &~   8.1  $\pm$   0.3  &~    4.5  $\pm$   0.5 \\
HD135153  &~   4.29  $\pm$   0.12  &~   4.48  $\pm$   0.09  &~   0.78  $\pm$   0.04  &~   4.96  $\pm$   0.16  &~   12.8  $\pm$   0.6   &~   8.7  $\pm$   0.3  &~    4.8  $\pm$   0.5 \\
HD164136  &~   2.30  $\pm$   0.08  &~   2.28  $\pm$   0.04  &~   1.10  $\pm$   0.03  &~   2.84  $\pm$   0.07  &~    7.4  $\pm$   0.3   &~   4.1  $\pm$   0.1  &~    3.7  $\pm$   0.3 \\
HD182835  &~   3.86  $\pm$   0.08  &~   3.80  $\pm$   0.06  &~   1.17  $\pm$   0.02  &~   4.35  $\pm$   0.26  &~   14.3  $\pm$   0.6   &~   8.1  $\pm$   0.3  &~    6.8  $\pm$   0.6 \\
HD213306  &~   0.95  $\pm$   0.10  &~   2.14  $\pm$   0.07  &~   1.50  $\pm$   0.02  &~   2.16  $\pm$   0.06  &~   14.0  $\pm$   0.4   &~   2.3  $\pm$   0.2  &~   12.0  $\pm$   0.4 \\
\hline
\end{longtable}
\end{center}

\begin{center}\small
\begin{longtable}{lrrrrrrrr}
\caption{EW of the $K$-band indices. Part 1.}
\label{tab:index_mis_K_1} \\

\hline
\hline
Star IDs  &  Fe1  &  Br$\delta$  &  Ca1  &  Fe23  &  Si  &  Ca2  &  Ca3  &  Ca4  \\
\hline                                                 
\endfirsthead
\multicolumn{9}{c}%
{{\bfseries \tablename\ \thetable{} -- continued from previous page}} \\
\hline
\hline
Star IDs  &  Fe1  &  Br$\delta$  &  Ca1  &  Fe23  &  Si  &  Ca2  &  Ca3  &  Ca4  \\
\hline                                                  
\endhead
\hline
\\
\multicolumn{9}{c}{\textbf{Supergiants}}\\
HD007927  &    0.30  $\pm$ 0.10  &    4.69  $\pm$ 0.20  &    0.48  $\pm$ 0.09  &    1.05  $\pm$ 0.14  &    1.17  $\pm$ 0.14  &    0.78  $\pm$ 0.05  &    0.87  $\pm$ 0.10  &    0.67  $\pm$ 0.07  \\
HD006130  &    0.60  $\pm$ 0.15  &    6.09  $\pm$ 0.30  &    0.81  $\pm$ 0.13  &    1.03  $\pm$ 0.21  &    1.35  $\pm$ 0.20  &    0.85  $\pm$ 0.08  &    0.82  $\pm$ 0.16  &    0.65  $\pm$ 0.11  \\
HD135153  &    2.03  $\pm$ 0.52  &    6.58  $\pm$ 1.04  &    1.18  $\pm$ 0.47  &    1.89  $\pm$ 0.76  &    1.05  $\pm$ 0.74  &    0.18  $\pm$ 0.28  & $-$0.09  $\pm$ 0.58  & $-$0.33  $\pm$ 0.42  \\
HD164136  &    0.67  $\pm$ 0.13  &    5.13  $\pm$ 0.25  &    0.80  $\pm$ 0.11  &    0.62  $\pm$ 0.18  &    0.69  $\pm$ 0.17  &    0.32  $\pm$ 0.06  &    0.15  $\pm$ 0.13  &    0.06  $\pm$ 0.10  \\
HD182835  &    1.12  $\pm$ 0.17  &    5.75  $\pm$ 0.33  &    0.72  $\pm$ 0.15  &    0.87  $\pm$ 0.24  &    0.59  $\pm$ 0.23  &    0.25  $\pm$ 0.09  & $-$0.13  $\pm$ 0.18  & $-$0.12  $\pm$ 0.13  \\
HD213306  &    1.99  $\pm$ 0.37  &    5.25  $\pm$ 0.74  &    1.86  $\pm$ 0.33  &    2.60  $\pm$ 0.53  &    1.93  $\pm$ 0.52  &    1.13  $\pm$ 0.20  &    1.05  $\pm$ 0.41  &    0.68  $\pm$ 0.29  \\
\hline
\end{longtable}
\end{center}

\begin{center}\small
\begin{longtable}{lrrrrrrrr}
\caption{EW of the $K$-band indices. Part 2.}
\label{tab:index_mis_K_2}  \\

\hline
\hline
Star IDs  &   Na$_{\rm d}$ &   Ca$_{\rm d}$  &   $^{12}$CO  &   Mg1  &   Br$\gamma$  &    FeA  &   FeB &   Mg2   \\
\hline                                                 
\endfirsthead
\multicolumn{9}{c}%
{{\bfseries \tablename\ \thetable{} -- continued from previous page}} \\
\hline
\hline
Star IDs  &   Na$_{\rm d}$ &   Ca$_{\rm d}$  &   $^{12}$CO  &   Mg1  &   Br$\gamma$  &    FeA  &   FeB &   Mg2   \\
\hline                                                  
\endhead
\hline
\\
\multicolumn{9}{c}{\textbf{Supergiants}}\\
HD007927  & $-$0.27  $\pm$ 0.04  & $-$0.18  $\pm$ 0.06  &    0.26  $\pm$ 0.15  & $-$0.05  $\pm$ 0.06  &    5.07  $\pm$ 0.01  & $-$0.06  $\pm$ 0.03  &    0.04  $\pm$ 0.03  & $-$0.11  $\pm$ 0.08  \\
HD006130  &    0.13  $\pm$ 0.05  & $-$0.14  $\pm$ 0.07  &    0.51  $\pm$ 0.18  &    0.05  $\pm$ 0.11  &    6.20  $\pm$ 0.05  &    0.01  $\pm$ 0.04  &    0.06  $\pm$ 0.05  &    0.10  $\pm$ 0.09  \\
HD135153  &    0.17  $\pm$ 0.10  & $-$0.06  $\pm$ 0.10  &    0.80  $\pm$ 0.24  &    0.04  $\pm$ 0.12  &    5.85  $\pm$ 0.05  & $-$0.13  $\pm$ 0.07  & $-$0.04  $\pm$ 0.08  &    0.17  $\pm$ 0.08  \\
HD164136  &    0.19  $\pm$ 0.06  & $-$0.03  $\pm$ 0.09  &    0.41  $\pm$ 0.21  &    0.07  $\pm$ 0.11  &    5.02  $\pm$ 0.05  & $-$0.00  $\pm$ 0.04  &    0.01  $\pm$ 0.04  &    0.26  $\pm$ 0.07  \\
HD182835  &    0.21  $\pm$ 0.06  & $-$0.18  $\pm$ 0.10  &    0.95  $\pm$ 0.25  &    0.05  $\pm$ 0.10  &    5.21  $\pm$ 0.05  & $-$0.05  $\pm$ 0.05  &    0.03  $\pm$ 0.05  &    0.17  $\pm$ 0.15  \\
HD213306  &    0.99  $\pm$ 0.08  &    0.40  $\pm$ 0.18  &    0.91  $\pm$ 0.44  &    0.63  $\pm$ 0.13  &    3.03  $\pm$ 0.10  &    0.19  $\pm$ 0.07  &    0.28  $\pm$ 0.08  &    0.67  $\pm$ 0.09  \\
\hline
\end{longtable}
\end{center}

\end{document}